\documentclass[prd,aps,onecolumn,a4paper,floatfix,nofootinbib,notitlepage]{revtex4-1}

\usepackage[utf8]{inputenc}
\usepackage{graphicx,psfrag}
\usepackage{mathrsfs}
\usepackage{amsmath,amsfonts,amssymb}
\usepackage{multirow}
\usepackage{comment}
\usepackage{xcolor}
\usepackage{enumerate}
\usepackage{hyperref}
\hypersetup{
    colorlinks = true,
    linkcolor = {blue},
    citecolor = {blue},
    urlcolor = {blue},
    linkbordercolor = {white},
    }

\begin{document}

\title{Efficient Kerr soliton comb generation in micro-resonator with interferometric back-coupling}

\author{J.M. \surname{Chavez Boggio}$^{1,*}$}
\author{D. \surname{Bodenmüller}$^1$} 
\author{S. \surname{Ahmed}$^{1}$}
\author{S. \surname{Wabnitz}$^2,^3$}
\author{D. \surname{Modotto}$^4$}
\author{T. \surname{Hansson}$^5$}

\affiliation{${}^1$ innoFSPEC-Leibniz Institut für Astrophysik, An der Sternwarte 16, 14482 Potsdam, Germany}
\affiliation{${}^2$ Dipartimento di Ingegneria dell'Informazione, Elettronica e Telecomunicazioni, Sapienza Universit\`a di Roma, via Eudossiana 18, 00184 Rome, Italy}
\affiliation{${}^3$ CNR-INO, Istituto Nazionale di Ottica, Via Campi Flegrei 34, 80078 Pozzuoli (NA), Italy}
\affiliation{${}^4$ Dipartimento di Ingegneria dell’Informazione, Università di Brescia, via Branze 38, 25123 Brescia, Italy}
\affiliation{${}^5$ Department of Physics, Chemistry and Biology,
Link\"oping University, SE-581 83 Link\"oping, Sweden}

\date{\today}

\begin{abstract}
Nonlinear Kerr micro-resonators have enabled fundamental breakthroughs in the understanding of dissipative solitons, as well as in their application to optical frequency comb generation. However, the conversion efficiency of the pump power into a soliton frequency comb typically remains below a few percent. We introduce a hybrid Mach-Zehnder ring resonator geometry, consisting of a micro-ring resonator embedded in an additional cavity with twice the optical path length of the ring. The resulting interferometric back coupling enables to achieve an unprecedented control of the pump depletion: pump-to-frequency comb conversion efficiencies of up to 98\% of the usable power is experimentally demonstrated with a soliton crystal comb. We  assess the robustness of the proposed on-chip geometry by generating a large variety of dissipative Kerr soliton combs, which require a lower amount of pump power to be accessed, when compared with an isolated micro-ring resonator with identical parameters. Micro-resonators with feedback enable accessing new regimes of coherent soliton comb generation, and are well suited for comb applications in astronomy, spectroscopy and telecommunications. 
\end{abstract}

\maketitle

\section*{Introduction}

State-of-the-art complementary metal-oxide semiconductor (CMOS) technology allows for the fabrication of arbitrary architectures with dimension accuracy approaching the nanometer level [1,2]. This has enabled the development of a variety of chip-based photonic devices with advanced functionalities. A relevant example is provided by optical frequency comb (OFC) generators based on integrated ring-resonators [3-7]. Fabricated by using dielectric or semiconductor materials, on-chip resonators are expected to down-scale the size of OFCs, offering miniature solutions for a number of applications ranging from telecommunications to metrology, astronomy and spectroscopy [8-15]. In its standard configuration, a bus waveguide is positioned in close proximity to a (ring) resonator, in order to couple pump light through its evanescent field [16-24]. Because of its simplicity, this geometry has a widespread use, and it has allowed breakthroughs in the generation of coherent OFCs based on dissipative Kerr solitons (DKSs) [25-32]. Nonlinear light propagation in Kerr micro-resonators is accurately modeled with the help of the so-called Lugiato-Lefever equation (LLE), which has permitted for a deeper understanding of the soliton generation dynamics [33-39].

A key limitation of present DKS-based micro-resonator OFC sources, though, is the extremely low conversion efficiency of the pump power into the power of the comb lines [40]. This is notoriously the case when generating bright solitons in the anomalous dispersion region of the ring waveguide [25]. The low transfer of pump power into the comb is partially due to a pump/resonator transverse mode mismatch. More importantly, since solitons are generated for an effective red detuning of the pump from micro-cavity resonances, most of the pump power is reflected, hence it is not coupled into the micro-resonator. Additionally, the pump-to-comb conversion efficiency is limited because of parametric gain saturation. For a single DKS comb, this results in comb lines carrying less than 1\% of the input pump power. Whereas for multi-solitons or soliton crystals, the comb lines to pump ratio is typically limited to be less than $\sim{5}$\%. Although the generation of dark soliton combs in the normal dispersion regime may bring the pump-to-comb conversion efficiency up to the 30\% range [20, 41], however dark solitons have a narrow spectral bandwidth, and exhibit a quite narrow domain of existence in the parameter space (power vs pump-resonance detuning) when compared with bright DKSs.

The development of a novel geometry and/or material platforms might enable the generation of frequency combs with much improved performances. In recent years, taking advantage of the high-accuracy of\, CMOS fabrication capabilities, a number of new architectures for generating OFCs have been proposed, mainly by incorporating the coupling of two adjacent resonators\, [42-48]. For example, it has been shown that, by placing two ring resonators in close proximity, a split of their resonances occurs, leading to the emergence of new DKS dynamics [44]. Interestingly, the use of two adjacent cavities enables nearly 100\% pump recycling, where one cavity stores the pump light, while a DKS is generated in the second cavity \,[45]. Here we introduce, fabricate and characterize a novel hybrid Mach-Zehnder micro-ring resonator architecture. It consists of a ring where DKSs are generated, embedded in a secondary cavity with twice optical path length, which acts as a feedback section. The interference of the fields from the ring and the feedback section enables for an unprecedented control of the pump depletion. This permits us to achieve almost a 100\% transfer of the pump power into the comb lines at the device output. The conversion efficiency is only limited by the pump power which is necessary to store in the ring, in order to sustain the DKS propagation. The robustness of the proposed architecture is effectively demonstrated by generating a large variety of DKSs, with usable conversion efficiencies up to 98\%. Our approach brings Kerr micro-combs from being inherently low-efficiency devices into comb generators possessing efficiencies which even surpass those of laser sources.

\section*{Results}

\subsection*{Ring resonator with optical feedback}

Figure 1(a) shows the schematic of the proposed device architecture for efficient DKS comb generation. A twisted bus waveguide couples pump light into a ring resonator through its evanescent field at two different coupling locations (marked as 1 and 2). The radius of the ring resonator is \emph{R}, while the nominal distance between the two evanescent coupling points is set to 3$\pi$\emph{R}. In this way, two coupled cavities are formed: the first cavity is the ring itself, while the second cavity comprises the twisted bus waveguide and the left semi-circumference of the same ring [49-52]. The power coupling coefficients ${\theta_1}$ and ${\theta_2}$ are adjusted by choosing different gaps between the bus waveguide and the ring at nodes 1 and 2, respectively. DKSs are generated in the ring section by injecting a red-detuned continuous wave (CW) pump. Therefore, a fraction of the pump power is coupled into the ring at node 1, acquiring a phase shift of $\pi$/2, while the remainder of the pump passes through. Furthermore, the DKS field generated in the ring that comes out at node 1, mixes with the pump field which is not coupled into the ring, creating the input field to the feedback section. At node 2, the feedback field and the field coming out from the ring mix, resulting in the device output field. Interestingly, the pump fraction that is in- and then out-coupled (through nodes 1 and 2), acquires a net phase shift of $\pi$. It is possible to obtain complete pump depletion, hence extreme conversion efficiency into the comb lines, if the pump fields coming out from the ring and from the feedback section have: i) equal amplitudes and ii) opposite phases. Figure 1(b) illustrates, in the frequency domain, the principle of the mechanism leading to complete pump depletion. After node 1, the frequency comb spectrum exhibits a strong pump component, as it occurs in an isolated micro-resonator, owing to the uncoupled pump field. On the other hand, inside the ring, the frequency comb spectrum exhibits a pump to comb-lines ratio which is smaller than in the feedback section. If the intra-cavity power is strong enough, it is possible that the out-coupled pump component at node 2 is equally intense as the pump component from the feedback loop. Whenever these pump fields are out-of-phase,
a strong pump depletion occurs at the output, resulting in extremely high conversion efficiency from pump to comb lines. 

\begin{figure*}[h]
\centering
\includegraphics[width=\linewidth]{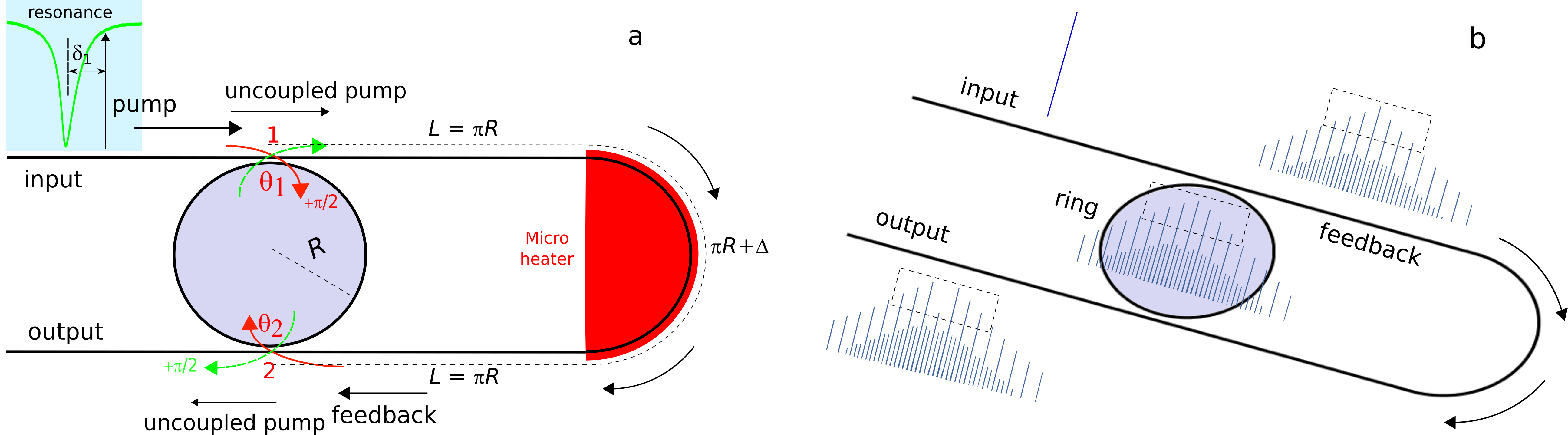}
\caption{Schematic of the ring resonator with an interferometric back-coupling architecture. (a) The ring is embedded in a cavity with twice its length. At the coupling points 1 and 2, there is a fraction of the field that is in- or out-coupled, while the remaining fraction is reflected. The interference of fields is adjusted by tuning (with a micro-heater) the length of the feedback section by ${\Delta}$. (b) Evolution of the frequency comb spectrum at the ring, feedback and output sections. If the pump fields coming out of the ring and feedback sections have similar amplitude and opposite phases, extreme pump depletion occurs at the device output.}
\label{figure:1}
\end{figure*}

\subsection*{Soliton generation with very high conversion efficiency}

The ring resonator with interferometric back-coupling was fabricated by using low-loss silicon nitride $({Si_X}{N_Y})$ with intrinsic \emph{Q}-factors exceeding ${10^6}$. Micro heaters on top of the feedback $Si_X N_Y$ waveguide allow for changing the optical path length of the outer cavity by a phase amount covering a ${2\pi}$ period. Figure 2(a) shows the experimental arrangement and the photography of the chip containing two ring resonators with feedback, used to generate DKSs. Both resonators have identical radius $R = 800\, \mu$m and first gap between bus waveguide and ring (at point 1) 650\, nm, whereas the second gap (at point 2) is  410 nm for resonator A and 460 nm for resonator B. The coupling coefficient at node 1 is ${\theta_1}\cong$ ${4.1\times10^{-3}}$ while at node 2 we have ${\theta_2}\cong$ ${3.77\times10^{-2}}$ (for resonator A) or ${\theta_2}\cong$ ${2.38\times10^{-2}}$ (for resonator B), which are obtained through full wave simulations (see Methods). The measured dispersion of the free-spectral range (FSR) in the resonator with feedback device is depicted in Figure 2(b). It was obtained by scanning the wavelength of a low-power laser between 1550 and 1630 nm (see supplementary note 2).

\begin{figure*}[h]
    \centering
\includegraphics[width=\linewidth]{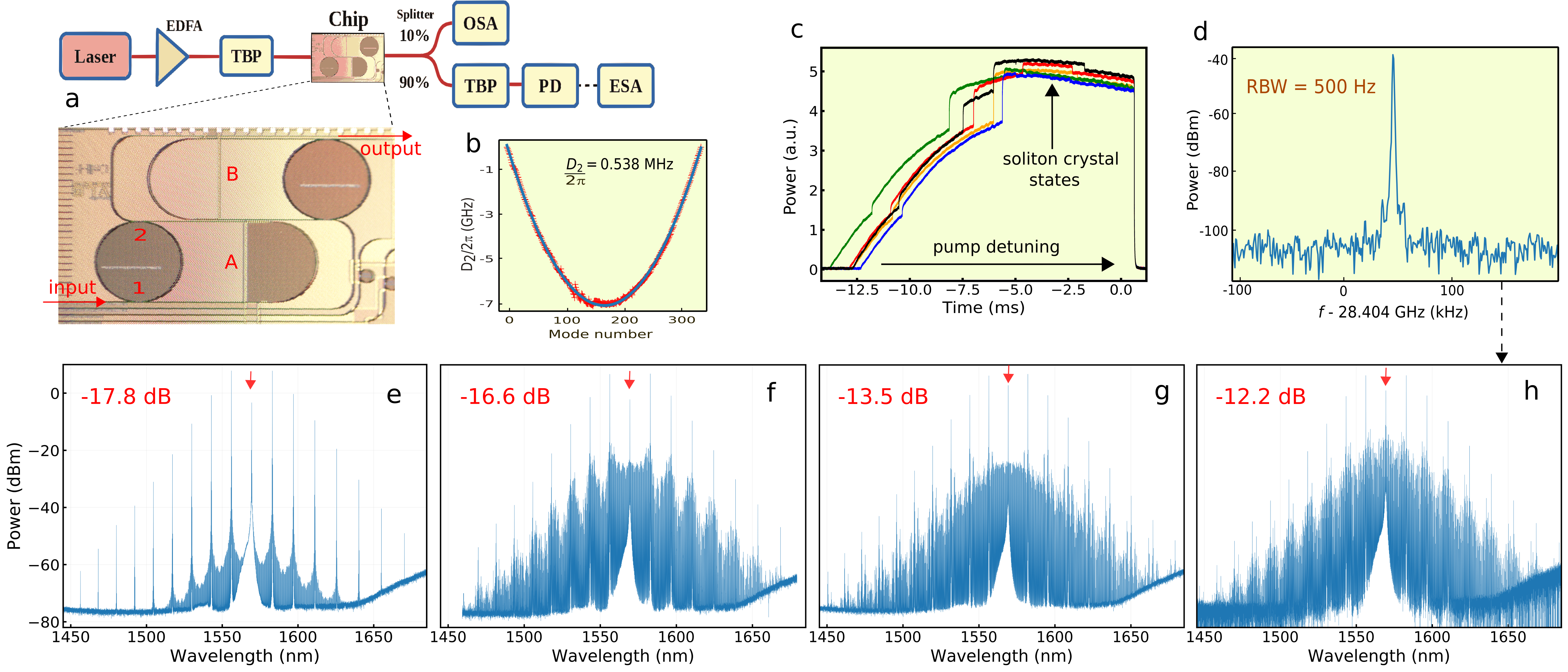}
\caption{ Efficient soliton generation in resonators with interferometric back-coupling. (a) Experimental setup and picture of a chip containing two ring resonators with feedback (A and B). TBP: tunable band-pass filter, PD: photodiode. OSA: Optical spectrum analyzer. ESA: Electrical spectrum analyzer. (b) Measured dispersion of the free spectral range featuring anomalous dispersion. (c) Output power of the generated comb as a function of the pump-resonator detuning. The different colors represent consecutive pump scanning through the resonance. (d) Repetition rate signal of the soliton crystal with defect shown in (h). Frequency comb spectra featuring a (e) perfect soliton crystal and (f-h) soliton crystals with defect. The pump depletion values indicated in red are measured with respect to the input power.}
\label{figure:2}
\end{figure*}

To generate the solitons, the quasi-TE mode of one resonance of the ring resonator with feedback A was pumped with a CW tunable laser, whose wavelength was tuned from the blue to the red. We pumped a resonance with measured loaded \emph{Q}-factor of ${0.55\times10^6}$ with a pump power inside the chip of 150 mW. Figure 2(c) shows the frequency comb output power from the device, versus pump-resonance detuning, obtained by consecutive scanning of the pump wavelength through the resonance. A Bragg filter was used to completely suppress the pump line. Several steps can be observed in the frequency comb power, which indicates the access to different comb regimes. Since the step heights change between different scans, this indicates that the access to the different comb regimes is done in a stochastic way. 

Figures 2(e-g) show typical spectra corresponding to the output power traces in Figure 2(c). For Figure 2(h) the pump power was increased to 180\, mW. Evenly spaced strong comb lines, accompanied by small comb lines with distinctive spectral features, indicate the generation of soliton crystals, as it occurs in isolated Kerr micro-resonators. However, the main difference that can be observed here, with respect to soliton crystals generated in Kerr resonators without feedback, is the strong pump depletion. The spectrum in (e) is a perfect soliton crystal (PSC) exhibiting a 17.8 dB depletion, i.e. only 1.7\% of the pump comes out from the device (i.e., micro-resonator with feedback). The pump-to-comb conversion efficiency can be defined as \emph{$\epsilon$} = ($P_{comb,out}$ / $P_{pump,in}$), where $P_{comb,out}$ is the output comb power excluding the power at the pump comb line, and $P_{pump,in}$ is the input pump power. By using this definition, the comb lines for the PSC in Figure 2(e) carry $\sim{55}$\% of the input pump power. The remaining $\sim{45}$\% of the pump power propagates in the ring, in order to sustain the DKSs. Nevertheless, the amount of pump power which is stored in the ring can be reduced to $\sim{25}$\% by increasing the coupling coefficients (see Supplementary note 3). In our resonator, the pump power depletion due to losses is $\sim{1}$\%. 

In Figure 2, and all along this work, soliton crystal states are distinguished from Turing rolls by exhibiting a red-detuned (as opposed to blue-detuned for rolls) pump wavelength: this is confirmed by measuring, with a counter-propagating probe, the pump-resonance wavelength detuning (see Methods). Furthermore, the repetition rate signal was measured for all spectra in Figures 2(f-h), showing the high coherence of soliton crystal with defect generation in resonators with feedback. Figure 2(d) shows the repetition rate for the spectrum in Figure 2(h).

It is interesting to compare soliton crystal generation in a resonator with feedback, with the soliton crystals obtained in experiments using isolated Kerr micro-resonators with a similar ring radius. In Figure 1(b) of [30], a PSC state was generated in a ring with a FSR of 20 GHz, corresponding to 87 solitons, evenly separated by 575 \,fs. In our case, the PSC in Figure 2(e) corresponds to a set of 58 pulses, evenly distributed along the ring circumference with 607\, fs spacing. The width of the solitons in Figure 2(e) is estimated from a Fourier transform calculation, by using 58 equidistant pulses. A pulse width of 110\, fs produces a PSC spectrum that matches well the experimental result. Even though the frequency bandwidth of the PSC in [30] is similar to that in Figure 2(e), conversion efficiencies are drastically different. Due to the absence of the interferometric back coupling in an isolated micro-ring, the pump is $\sim{17}$ dB more intense than its immediate comb sidebands. Whereas, in our case the pump is 11\, dB less intense than its nearby comb lines.

\begin{figure}[hbt!]
\centering
\includegraphics[width=0.85\linewidth]{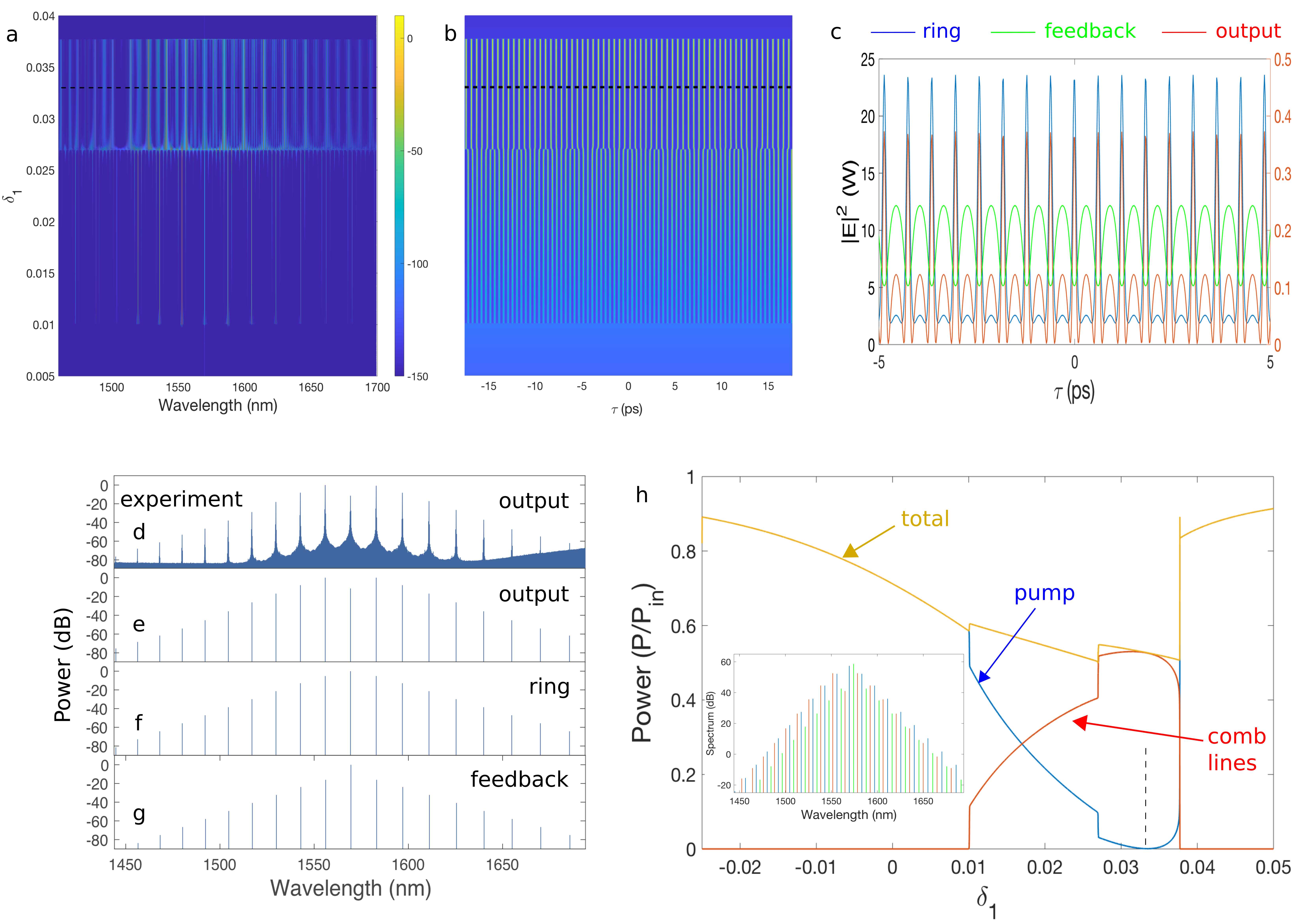}
\caption{Ikeda map simulations. Evolution as a function of the pump detuning ${\delta_1}$ of (a) intra-cavity frequency comb spectrum, and (b) intra-cavity temporal intensity. (c) Snapshot of the pulses at ${\delta_1}$ = 0.033 for ring (blue), feedback (green), and output (red) sections. The left y-axis corresponds to the ring power, while the right y-axis to the feedback and output powers. Snapshot of the frequency comb spectrum calculated at the (e) output, (f) ring, and (g) feedback sections of the device. Panel (d) shows the corresponding experimental output spectrum. (h) Calculated power carried by the pump (blue), comb lines (red), and entire comb (dark yellow) vs. detuning ${\delta_1}$. Note that complete pump depletion is obtained for ${\delta_1}$ between 0.03 and 0.035. The inset of (h) shows the comparison of the comb spectra at the ring (re-scaled, blue), feedback (green) and output (red) sections.}
\label{figure:3}
\end{figure}

In order to model nonlinear pulse propagation in a micro-resonator with feedback, we use an Ikeda map with a delay, associated with the field from the feedback arm. For the Ikeda map simulations we use the parameters of the ring resonator with feedback that are obtained through measurements and numerical calculations (see Methods). Figure 3(a) shows the evolution of the frequency spectrum in the micro-ring, while Figure 3(b) shows the associated temporal evolution, as the detuning ${\delta_1}$ is varied in a resonator with feedback with the same parameters as in Figure 2, and 180 mW of pump power. The detuning offset of the feedback arm is set to ${\delta_{20}}$ = 1.5 (see suplementary note 1). Note that there is first the formation of Turing rolls at ${\delta_1}$ $\sim{0.01}$, followed by the merging of those pulses, and the appearance at ${\delta_1}$ $\sim{0.027}$ of a train of pulses with larger inter-soliton separation. By increasing further the pump detuning, the pulses vanish at ${\delta_1}$ $\sim{0.038}$. Figure 3(c) compares the temporal traces at the ring (blue), feedback (green), and output (red) sections, for a detuning of ${\delta_1}$ = 0.033, where the strongest pump depletion occurs at the device output. The pulses propagating in the ring are sitting on a weak background, while the pulses in the feedback section feature a dark pulse shape. On the other hand, the output pulses exhibit a strong contrast with the background. We also show the corresponding frequency comb spectra at the three sections: output (Figure 3(e)), ring (Figure 3(f)), and feedback (Figure (g)), respectively. Although all spectra look similar, their main difference is found in the intensity of the continuous wave pump component. The pump is around 17 dB stronger than its immediate sidebands in the feedback section, while it is only around 7 dB stronger in the ring section. At the device output, the pump is depleted by as much as 25 dB. For a comparison, Figure 3(d) depicts the experimental spectrum, showing an excellent agreement with the simulation. To confirm that the pulses generated at ${\delta_1}$ = 0.033 indeed correspond to DKSs, we numerically verified that they continue to propagate as single solitons after the removal of all but one pulse (see Supplementary note 3).

Figure 3(h) shows the evolution, as a function of detuning ${\delta_1}$, of the output power at the pump component (blue), comb lines (red), and for the entire comb (dark yellow). There is a range of pump detunings, between ${\delta_1}$ = 0.03 and 0.035, where the pump is completely depleted, and its power is transferred into the comb lines. The fact that the total output power drops to 55\% of the input power at maximum depletion, is due to the pump power which remains circulating in the ring in order to sustain the DKSs. Interestingly, we may obtain physical insight into the conditions for strong pump depletion to happen, by taking the calculated spectrum in the ring section and multiplying it by the power coupling coefficient ${\theta_2}$ = ${3.77\times10^{-2}}$. In this way, we estimate the field coming out from the ring at node 2, and we find that, for the pump component, it has almost the same intensity as that emerging from the feedback arm. This can be seen in the inset of Figure 3(h), where the three spectra (re-scaled ring, feedback and output) are shifted in wavelength, for better clarity. Note that, contrary to the pump, the comb lines from the (re-scaled) ring and feedback sections have very dissimilar intensities.

\subsection*{Generated comb power, pump detuning, and access to diverse soliton states}

Soliton dynamics in isolated micro-ring resonators are governed by two parameters: the intra-cavity power and the effective detuning 
[53]. Experimentally, the pump power is kept constant and the detuning is varied to access DKSs: the route that the pump undergoes in the 
parameter space determines which particular DKS or chaotic state is going to be accessed. A convenient way to compare Kerr resonators 
having different parameters (such as FSR, nonlinearity, Q-factor, etc...) is obtained by using a normalized pump amplitude \emph{f}, 
where $f^2 = \dfrac{8g\eta}{\Delta \omega_{T}^2} \dfrac{P}{\hslash \omega_0}$, with $P$ being the input pump power, ${\omega_0}$ 
is the resonance frequency, ${\hslash}$ is the reduced Planck constant, ${\Delta \omega_{T} = \Delta \omega_0 + \Delta \omega_{ext}}$ is 
the total cavity linewidth, which is the sum of the intrinsic linewidth ${\Delta \omega_0}$ and the coupling linewidth ${\Delta 
\omega_{ext}}$. The coupling efficiency is given by ${\eta = \Delta \omega_{ext}/\Delta \omega}$\,. The nonlinear gain, which accounts 
for the Kerr frequency shift per photon, is ${g = (\hslash \omega_0^2cn_2)/(n^2 V_{eff})}$ (see Methods for details).

 It has been shown that by scanning the pump wavelength from blue to red, perfect soliton crystals\, are deterministically generated in micro-resonators at relatively low pump powers, and up to a normalized pump amplitude of $\emph{f}\sim{3}$\, [30]. At $\emph{f}\sim{3}$\, a spatiotemporal chaos\, (STC) regime is reached, and a PSC cannot be any longer generated with 100\% probability: soliton crystals with defects start to be generated in a stochastic manner. Furthermore, at a normalized pump amplitude of $\emph{f}\sim{4}$\, the transient chaos (TC) regime is reached, and PSCs and soliton crystals with defect can not be any longer accessed and only multi-soliton states are accessed by scanning the pump detuning [30]. To compare DKS generation in isolated micro-resonators with that occurring in a resonator with interferometric back-coupling, we show in Figure 4 traces of the frequency comb output power, versus pump-resonance detuning, obtained for different pump input powers in resonator B. In those measurements the pump line was removed with a Bragg grating. The relative phase between the cavities was not optimized to enhance the pump depletion and for this case the measured loaded \emph{Q}-factor is ${1.6\times10^6}$ at the pumped resonance. Note that for pump powers of 50\, and 68\, mW, the traces in Figure 4 exhibit soliton crystal steps at the highest frequency comb output power values: the steps' slope is positive, hence it is thermally stable and very easy to access. Successive scannings result in very minor changes in the output power traces, indicating that a deterministic access to the PSC states occurs, similar to the case of isolated Kerr resonators. On the other hand, for larger pump powers (i.e., for 80\, and 140\, mW), different scannings result in frequency comb power traces with soliton steps of different heights, which indicates that the STC regime has been reached. For the 80\, mW case for example (Figure 4(c)), two types of steps can be noticed: i) soliton crystal steps which occur with a small power drop and ii) multi-soliton steps that exhibit a much larger power drop. The upper plot of Figure 4(c) shows the spectrum of a soliton crystal with defect, that is generated with 80 mW of pump power. For a pump power of 140\, mW, as it is shown in Figure 4(d), no soliton crystal steps are observed, and only multi-soliton steps are present, which indicates that the TC regime has been reached. Since different step levels indicate the presence of different soliton numbers, from the results in Figure 4 we may conclude that a large variety of DKS states can be accessed in the resonator-with-feedback architecture.

\begin{figure*}[hbt!]
\centering
\includegraphics[width=\linewidth]{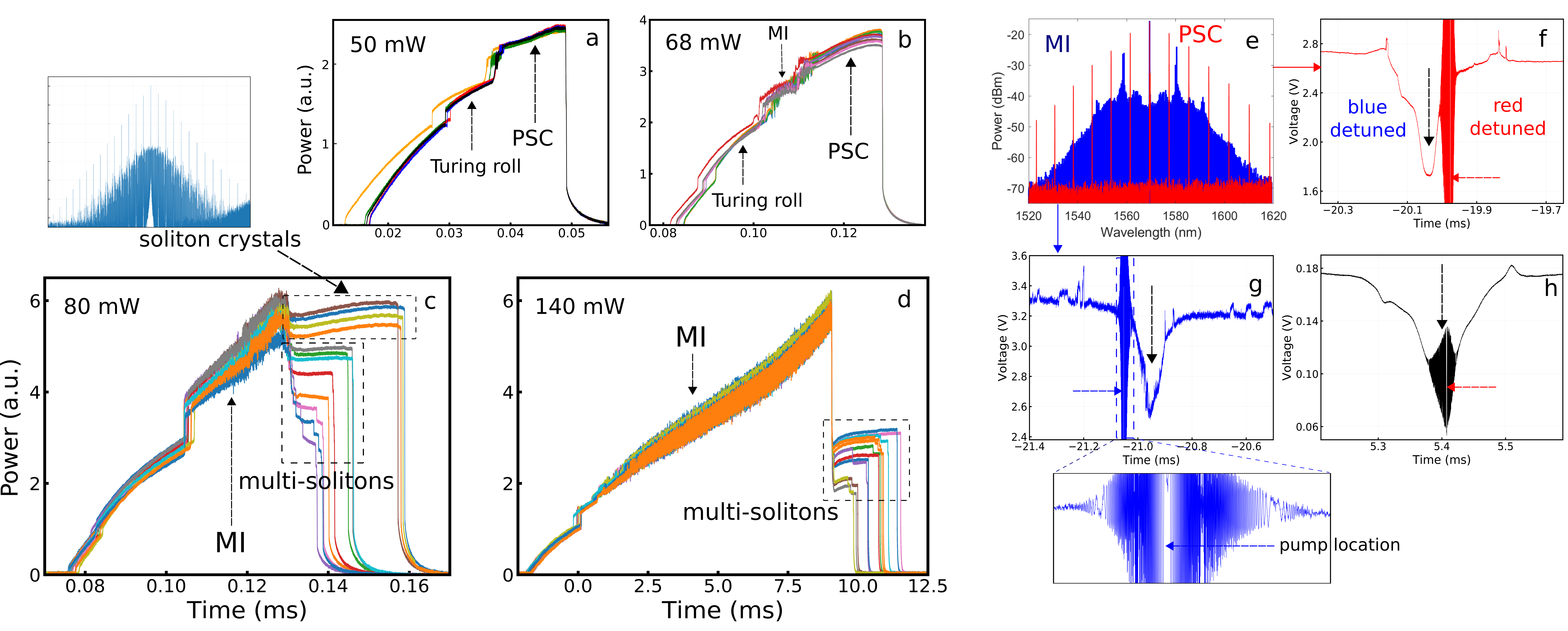}
\caption{Output power of the generated comb as a function of pump-resonance detuning in a resonator-with-feedback, for different input  pump powers. (a) 50 mW, (b) 68 mW, (c) 80 mW, and (d) 140 mW. Perfect soliton crystals steps are observed for 50, 68 and 80 mW. Multi-soliton steps are observed at 80 and 140 mW. For 80 and 140 mW pump powers, the step height changes for different scannings. The soliton crystal with defect spectrum shown above Figure (c) exhibits a typical spectrum generated at 80 mW, for the detuning that is indicated with a black arrow. (e) MI (blue line) and PSC (red line) spectra. Measurement of the effective pump detuning for the (f) PSC and (g) MI combs. The black arrow in (f), (g), and (h) indicates the center of the resonance. The panel below (g) shows the details of the beating between the pump and the probe. The case in (h) shows the smallest detuning that we could measure in a PSC.}
\label{figure:4}
\end{figure*}

Interestingly, soliton crystals with defect were even obtained for a pump power value of $\emph{P}\sim{70}$\, mW, i.e., with $\emph{f}\sim{1.5}$\,. This corresponds to a pump power four times smaller than the value for an isolated micro-resonator (see Methods for a more detailed description). Errors in determining \emph{f} might come from an imperfect knowledge of the nonlinear refractive index, and mainly from not knowing the exact pump power that propagates inside the bus waveguide (1\, dB error). Even taking into account those error sources, the STC regime starts at $\emph{f}\sim{2}$\,. Furthermore, the TC regime has its lower boundary at $\emph{f}\sim{2.8}$\, instead of $\emph{f}\sim{4}$\,. Therefore, in Kerr resonators with feedback some DKS states can be accessed by tuning the pump wavelength from blue to red by using a lower amount of pump power than in an isolated Kerr resonator. This can be related with the pump power re-use, that in our case occurs because of the feedback. Whereas in the case of an isolated micro-resonator the pump that is not coupled into the ring does not contribute to the parametric gain process.   

In isolated micro-resonators, DKSs are generated for an effective red-shifted detuning of the pump. For the case of single solitons, the detuning can be many times larger than the resonance linewidth, which results in a very poor conversion efficiency from the pump into comb lines. We investigated the effective pump detuning while generating frequency combs by injecting a weak counter-propagating probe, whose wavelength scans the resonance at a 10 Hz rate [21]. The power of the probe is adjusted, in order to match the power of the pump that is back-reflected, owing to imperfections in the resonator. Light from the probe and the reflected pump is filtered with a band-pass filter, in order to reject the comb lines, and it is detected with a fast photodiode. In this way, not only the resonance trace is obtained, but the interference pattern which is produced by the beat between the moving probe and the fixed pump is also recorded. The exact location of the pump corresponds to the position where the beat note frequency vanishes. Figure 4(e) shows in a red color the spectrum of a perfect soliton crystal, and in blue we illustrate the case of a chaotic modulation instability (MI) state. The spectra correspond to the same resonator and power as in the case of Figure 4(c). Figures 4(f) and 4(g) show the measurement of the effective detuning for the PSC and MI spectra, respectively. Similar to the case of a single micro-resonator, bistability results in the distinction between chaotic states, which are accessed with a blue-detuned pump wavelength, and soliton states, which are obtained with a red-detuned pump. The panel below Figure 4(g) shows the detail of the interference between the pump and the probe lasers. The pump location with respect to the resonance can be accurately measured by retrieving the zero of the beat note frequency.

Soliton crystals (perfect and with defect) were generated with different levels of conversion efficiency, or pump depletion, and their corresponding pump detuning was measured. For all cases, the red detuning was not bigger than $\sim$\,1.5 times the resonance linewidth. However, no clear correlation between the value of detuning and the associated level of pump depletion was observed, even though we would expect that a smaller detuning could be associated with stronger pump depletion. Figure 4(h) shows an example of the detuning measurement of a soliton crystal for which the red detuning is the smallest we have observed.

\begin{figure*}[h]
\centering
\includegraphics[width=\linewidth]{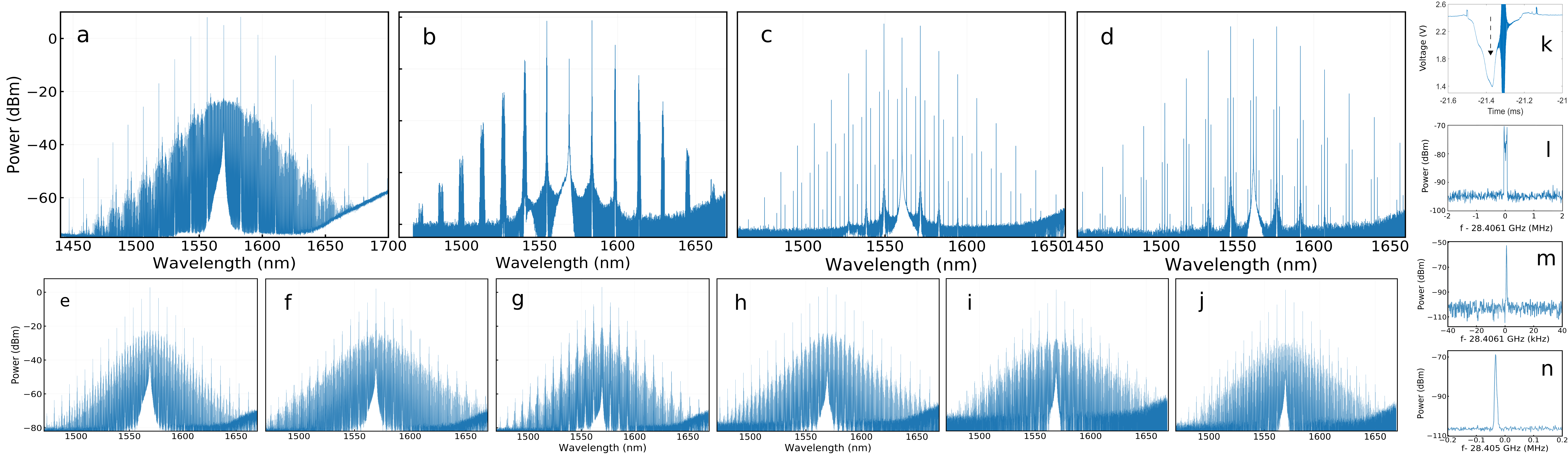}
\caption{Coherent frequency comb spectra featuring typical soliton crystal arrangements of strong and weak comb lines. The spectra of soliton crystals were obtained with (a-c) resonator A, (d-j) resonator B. The spectrum in (b) was obtained for a pump power of 300 mW and exhibits a strong enhancement of the comb lines. The spectra of soliton crystals with defect in (e-j) exhibit distinctive patterns of intensity variations of the small comb lines. (k) depicts the effective pump detuning measurement corresponding to the spectrum in (c). (l-m) repetition rate signals of (b), (i), (j), respectively.}
\label{figure:5}
\end{figure*}

The complexity of adding a feedback arm to the micro-ring resonator does not seem to drive the device into a generator of chaotic states, but rather into a generator of stable DKSs. Figure 5 shows the variety of coherent and red detuned frequency comb spectra that can be generated in the ring resonator with feedback. Figures 5(a-c) show frequency comb spectra obtained with resonator with feedback A, from the same set of measurements of Figure 2 but varying the pump power. The case of Figure 5(b) shows an OFC exhibiting an extreme pump depletion, 22 dB with respect to the input power, i.e. 99.4\% of the pump power is converted into the comb lines at the output or circulates in the ring to sustain the comb generation. The Figures 5(d-j) show frequency combs generated in the resonator B. For the case of Figure 5(d) the relative phase between the cavities was adjusted with the micro-heater to enhance the pump depletion. The spectra in Figures 5(e-j) correspond to the frequency comb power traces shown in Figure 4(c). In those spectra we depict different soliton crystals with defect that were generated by pumping with power values of 70-100\, mW. The interference between strong and weak comb lines results into a set of DKSs that are distributed along the resonator circumference, where some pulses are expected to be missing with respect to the perfect SC. For the comb spectra in Figures 5(a,h), we calculated their inverse Fourier transform by taking into account that all strong lines are in phase, while the weak lines are out of phase by $\pi$ with respect to the strong ones [26, 29].

Although an infinite range of defects and shifts in the soliton locations are possible, results that have been reported over the last couple of years by different groups show that some soliton crystals are more likely to be generated than others. For example, the OFCs in Figures 5(c,d,e,i,j) were previously reported by using different architectures and materials [26,29,30,32]. The advantage of using a micro-resonator with the extended cavity architecture comes from the much higher conversion efficiency that is available at its output, when compared with the case of a single micro-resonator, as seen in Figures 5(c-d).

All the spectra exhibited high coherence of the repetition rate signal and red detuned pumps. Figure 5(k) shows the pump detuning measurement of the spectrum in Figure 5(c), where a red pump detuning can be observed. Figures 5(l, m, n) show the repetition rate signals, measured with a resolution bandwidth of 1 kHz, for the spectra in Figures 5(b, i, j), respectively. 

\subsection*{Effect on the conversion efficiency by tuning of $\Delta $}

The strong pump depletion which is obtained with our microcomb architecture originates from the interference at node 2 of the pump fields coming out from the ring and from the feedback section. The adjustment of the interference process relies on the knowledge of the phase mismatch between the cavities. We retrieve this mismatch, by measuring the linewidth of the resonances, since this linewidth depends on the feedback length, ${\Delta}$, which is tuned with the micro heater (see Supplementary note 3) [49-52]. The resonances linewidth in our device is approximately given by

 \begin{equation}
  \begin{aligned}
\Delta \omega_{T} \cong [\alpha L_1 + (\theta_1 + \theta_2)/2  + \sqrt{\theta_1\theta_2} \cos{(\beta_0 \Delta)}]/ \beta_1 L_1,
 \end{aligned}
\end{equation}

\noindent where $\alpha$ is the absorption loss, $L_1 = \pi R$, ${\beta_0} = 2 \pi n_{eff}/ {\lambda}$ is the propagation constant, and 
$\beta_1 = {t_R/2L_1}$ is the inverse of the group velocity, with ${t_R}$ being the round-trip time. Experimentally, we change ${\Delta}$ 
up to a corresponding phase shift slightly larger than $2\pi$, by delivering an electrical power into the micro-heaters. The linewidths 
and depths of the resonances are measured by performing a scan from 1550 to 1630\,nm with a tunable laser. From the 330 measured 
linewidths, we extract an average loaded ${\emph{Q} = {\omega_0} / {\Delta \omega_{T}} }$ for each value of ${\Delta}$. We performed the 
measurement for the two contiguous ring resonators with feedback. The result for resonator B is shown in Figure 6(a). The blue squares 
show the measured average \emph{Q}-factors, while the black solid line shows the calculation using Equation (1), which we plot as a 
function of the detuning of the feedback cavity ${\delta_{20} \cong {-\beta_0 \Delta}}$. The agreement between experiment and analytical 
calculation is very good, even though no fitting parameters were used. By correlating the measurement of the loaded average 
\emph{Q}-factor with the analytical calculation, it is possible to retrieve the value of ${\Delta}$. With red dashed lines we indicate 
the conversion efficiency values of perfect soliton crystals that were generated in the resonator with feedback B by using a constant 
pump power. We define the conversion efficiency as the ratio between the power of the comb lines to the total output power, i.e., we 
exclude the power which is stored in the ring in order to sustain the solitons. Even though high conversion efficiency is obtained over 
all ${\delta_{20}} $ values, nevertheless the highest conversion efficiency values are obtained around the lowest loaded 
\emph{Q}-factors, which correspond to a $\delta_{20}$ around 0. Figure 6 (b) shows the measured \emph{Q}-factor abundance corresponding 
to ${\delta_{20} = {\pi}}$. Figure 6(c) depicts contour plots of the intra-cavity power for different values of the phase mismatch 
between the cavities, ${\delta_{20}}$, and the detuning ${\delta_1}$. It can be noted that there are two resonances for the intra-cavity 
power at ${\delta_{20}}$ values around $ {\pi}$.

\begin{figure*}[h]
    \centering
\includegraphics[width=0.75\linewidth]{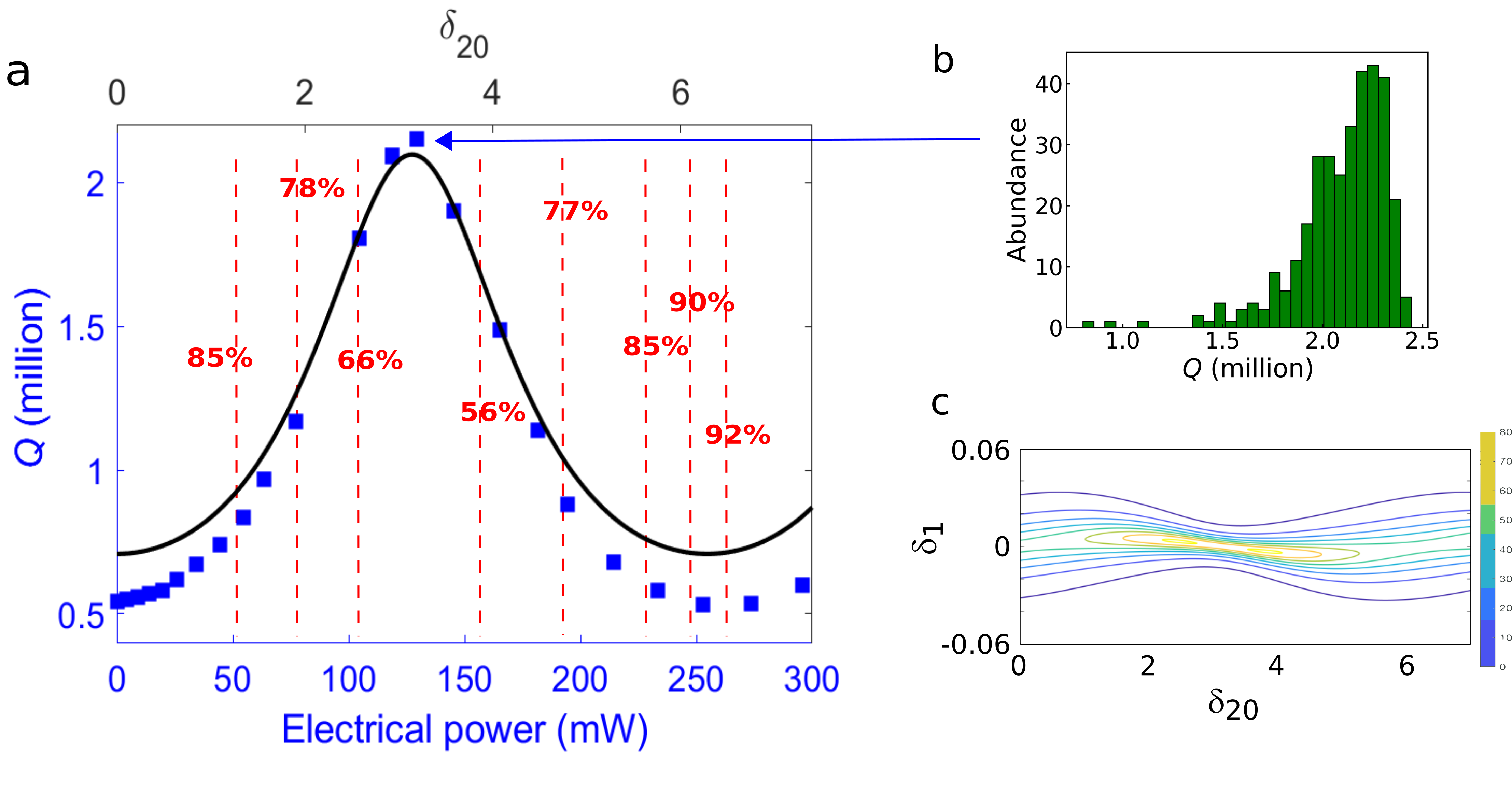}
\caption{Interferometric tuning of the conversion efficiency. ${\Delta}$\, (${\delta_{20}}$) is changed by applying electrical power on a thermal heater. (a) The  measured average \emph{Q}-factor as a function of the electrical power is depicted in squares while the calculated values from equation (1) are shown in a solid line. (b) loaded \emph{Q}-factor abundance for a relative phase between the cavities which is indicated by the blue arrow. (c) Contour plot of normalized intra-cavity power $I_B/I_{in}$ vs ${\delta_1},{\delta_{20}}$ in the linear limit.}
\label{figure:6}
\end{figure*}

Being a hybrid Mach-Zehnder interferometer ring resonator, the transmission function of our device does not depend on the direction of pump light injection, i.e., by inverting the propagation direction one obtains the same results. Even though the parametric dynamics is complex in our architecture, a pump phase examination can be performed. Depending on the pump wavelength detuning, there is a fraction of the pump that is never coupled into the ring (either in node 1 or 2) and mixes with the pump field that is in- and out-coupled from the ring. They will have a phase difference of $\pi$, favouring a pump depletion at ${\Delta}$\, around 0. Nevertheless, to obtain high pump depletion, the amount of photons that are never coupled into the ring cannot be excessively high, i.e. the pump detuning cannot reach excessively high values.

\section*{Discussion}

Embedding a micro-ring in a large fiber cavity which incorporates an optical amplifier has been investigated in the past years, enabling the recent demonstration of a laser cavity soliton comb [54-56]. Furthermore, supercontinuum generation in photonic crystal fibers having a fiber feedback section showed noise reduction, for some feedback conditions [57-58]. However, the on-chip hybrid Mach-Zehnder ring resonator geometry has not been yet investigated, to the best of our knowledge, for nonlinear propagation and OFC generation. This architecture was fabricated by using CMOS-based low-loss silicon nitride, enabling to accurately design the length of the two cavities to enhance the interferometric effect. We demonstrate that on-chip resonators with optical feedback support the generation of a rich variety of DKSs, while requiring a lower amount of pump power to be accessed when compared with a single micro-ring resonator architecture. Interferometric back-coupling allows for a strong pump-to-comb energy transfer, that results in a conversion efficiency of about 98\% of the usable power within a perfect soliton crystal. The resonator with feedback geometry shows that the transformation of a CW pump into a train of solitons can be done in a very robust manner, which is not destroyed by the feedback, but reinforced by it. This enables the access of soliton regimes which are not supported by isolated Kerr resonators. The ultra-efficient perfect soliton crystals reported in this work are particularly suitable for telecommunication applications. We envision that a variety of other applications could strongly benefit from the rich variety of coherent soliton combs that can be accessed with this geometry, such as in astronomy and spectroscopy. Furthermore, the generation of dissipative solitons in cavities with a interferometric back-coupling offers a new platform for studies of complex nonlinear light dynamics.  

\section*{Acknowledgments}

This work was supported by the BMBF (Federal Ministry of Education and Research) through grants 03Z2AN11 and 03Z2AN12. The work of S.W. was supported by the European Research Council (ERC) under the European Union’s Horizon 2020 research and innovation program (grant No. 740355 and grant No. 874596). T.H. acknowledges funding from the Swedish Research Council (Vetenskapsrådet, Grant. No. 2017-05309).

\section*{Methods}

\subsection*{Experimental set-up details} The light of the tunable laser is amplified with an Erbium-doped fiber amplifier\, (EDFA); a polarization controller permits to adjust the input pump polarization, in order to excite the quasi-TE mode of the micro-ring resonator. We used an objective lens\, (or a lensed fiber in some cases) to couple pump light into the chip, which contains inverse tapered bus-waveguides to minimize losses during the in- and out-coupling process (${\sim2.5}$\, dB at each coupling end). The generated frequency comb is visualized by using an optical spectrum analyzer (OSA). To measure the repetition rate of the frequency comb, we used a tunable Bragg grating to filter out the pump light, and we detected the comb light power with a fast photodiode, which permits its visualization with an electrical spectrum analyzer\, (ESA). The stability of the repetition rate signal is characterized by mixing it with a high-purity signal at 28.5\, GHz, and by measuring the down-converted signal with a frequency counter at different gate times. Both the signal generator and the frequency counter are disciplined with a Rubidium clock. Diagnostic tools include also a fast oscilloscope to detect the generated frequency comb power. Once a frequency comb is generated, we measure the effective pump-resonance detuning by counter-propagating a probe laser whose wavelength is scanned through the resonance with a zigzag function at 10\, Hz rate.

\subsection*{Parameters of the ring resonator with feedback}

The measurements reported in the main text were performed by using two nominally identical chips (1 and 2). In chip 1, wire bonding was added to provide electrical power to the micro heaters, and to adjust the relative phase between the cavities. Whereas in chip 2 no wire bonding was added. The chip contains two resonators with feedback, having nominal parameters $\emph{R} = 800~\mu$m, gap1 = 650 nm, while the gap2 = 410 (resonator A), or 460\, nm (resonator B). The core height and core width were the same for both resonators: 825\, nm and ${1.5~\mu}$m, respectively. This ensured anomalous chromatic dispersion at 1560\, nm. Besides the two resonators with feedback, the chip contains standard resonators that were fabricated in order to better retrieve the intrinsic losses and coupling coefficients of the resonators with feedback. The intrinsic \emph{Q}-factor in our resonator with feedback is estimated to be ${3.5\times10^{6}}$.

Numerical simulations were performed using RSoft FullWAVE to retrieve the power coupling coefficients between the ring resonator and the 
bus waveguide, by using the refractive index and cross-section values of the resonator. For the three gap values in our resonators, i.e. 
650\, nm, 460\, nm, and 410\, nm we obtained\, ${\theta_1}\cong$ ${4.1\times10^{-3}}$, ${\theta_2}\cong$ ${2.38\times10^{-2}}$, and\, 
${\theta_2}\cong$ ${3.77\times10^{-2}}$, respectively. All values were calculated at 1560\, nm. To assess the correctness of these 
calculations we did the following: we measured the coupling coefficient of the single resonators fabricated in the same chip with various 
gaps, and by comparing the measured with calculated values, a good agreement was confirmed. In a single resonator with our parameters, 
critical coupling would be obtained at a gap between the bus waveguide and the resonator of 600\, nm. The intrinsic losses are estimated 
to be 0.1\, dB/cm, therefore the fractional transmitted power along the ring and feedback sections are ${t_1}$ = 0.997 and ${t_2}$ = 
0.9914, respectively. The ${Si_{X}N_{Y}}$ resonators have linear and nonlinear refractive indices of $n$ = 2.1\, and\, ${n_{2} \cong 
2.4\times10^{-19}}$~m$^2$/W, respectively. The effective area is calculated to be ${A_{\rm eff} = 0.98~\mu m^{2}}$, while the effective 
volume $V_{\rm eff} = 2\pi R A_{\rm eff} = 4.92 \times 10^{-15}$~m$^3$ for $\emph{R} = 800$~$\mu$m. In our ${Si_X}{N_Y}$ resonators with 
feedback used in the experiments reported in the main text, we have the nonlinear gain coefficient $g = (\hslash \omega_0^2 c n_2)/(n^2 
V_{eff}) \cong 0.5$~Hz.

For the frequency comb output power traces in Figure 4 we pumped a resonance at 1569.3 nm that has\, ${\Delta \omega/2\pi \cong 120}$\, MHz, therefore ${\eta = \Delta \omega_{ext}/\Delta \omega}$\,${ \cong 0.55}$. By pumping with ${P = 70}$\, mW, i.e. a normalized pump amplitude of\, ${f \cong 1.5}$, the generation of soliton crystals with defect was observed. This corresponds to a pump power 4 times smaller than what would be expected in a single resonator without feedback. 

\subsection*{Ikeda map}
To model nonlinear propagation in a microresonator with feedback, we use an Ikeda map with a temporal delay for the field propagating through the feedback arm. The fields are assumed to have durations equal to the roundtrip time ${t_R}$ of the microring,
and to continuously move with the group-velocity  $\beta_1 = {t_R/2L_1}$ of the pump frequency, where ${2L_1}$ is the ring circumference.
At coupling node 1, the input field ${A_{in}}$ is split in two parts that propagate through the microring ${A}$ and the feedback arm ${C}$. This is modelled by the coupling conditions
\begin{equation}
 \begin{aligned}
{A_{m+1}(0,\tau)} = i{\sqrt{\theta_1}}A_{in} + {\sqrt{1 - \theta_1}}e^{-i\delta_{1}}B_{m}(L_{1},\tau), \qquad C_{m+1}(0,\tau) = {\sqrt{1 - \theta_1}}A_{in} + {i\sqrt{\theta_1}}e^{-i\delta_{1}}B_{m}(L_{1},\tau),
\end{aligned}
\end{equation}
where subscript ${m}$ denotes the roundtrip number, and ${2\delta_{1}}$ is the detuning of the ring cavity. The subsequent evolution in each waveguide is modelled by nonlinear Schrödinger equations
\begin{equation}
\dfrac{\partial A_{m}}{\partial z} = \left[-\dfrac {\alpha_i}{2}-i\dfrac{\beta_2}{2} \dfrac{\partial^2}{\partial \tau^2}\right] A_m + i \gamma|A_m|^2 A_m, \qquad \dfrac{\partial C_{m}}{\partial z} = \left[-\dfrac {\alpha_i} {2}-\Delta \beta_{1}\dfrac {\partial}{\partial \tau}-i\dfrac {\beta_{2}}{2} \dfrac {\partial ^{2}}{\partial \tau^{2}}\right]C_{m} + i \gamma|C_{m}|^{2} C_{m},
\end{equation}
where $\alpha_i$ is the absorption loss, $\beta_2$ is the second-order group-velocity dispersion and
${\gamma}$ = ${n_2\omega_0}/{(c A_{eff})}$ is the nonlinear coefficient. The fields ${A}$ and ${C}$ are propagated through distances ${L_1}$ and ${L_2}$ = 3${L_1}$, and advanced in time by ${t_R /2}$ and
3${t_R /2}$, respectively. Since there is a one round trip delay in the time required for the feedback field to reach node 2,
the coupling at node 2 occurs with the (stored) feedback field from the previous roundtrip
\begin{equation}
 {B_{m}(0,\tau)} =  \sqrt{1 - \theta_2} e^{-i\delta_{1}} A_{m}(L_{1},\tau) + i{\sqrt{\theta_2}}e^{-i\delta_{2}}C_{m-1}(L_{2},\tau),
\end{equation}
where ${\delta_{2}}$ is the detuning of the feedback section. The final evolution of the field ${B}$ from node 2 back to node 1 over the distance ${L_1}$ and with the temporal advance ${t_R /2}$ is again modelled by a nonlinear Schrödinger equation
\begin{equation}
 \dfrac{\partial B_{m}}{\partial z} = \left[-\dfrac {\alpha_i} {2}-i\dfrac {\beta_{2}}{2} \dfrac {\partial ^{2}}{\partial \tau^{2}}\right]B_{m} + i \gamma|B_{m}|^{2} B_{m}.
\end{equation}
To account for small differences in length between the two paths 3${L_1}$ (${A}$ → ${B}$ → ${A}$ → ${B}$) and ${L_2}$ (${C}$ → ${B}$), a group-velocity mismatch term is included in the evolution equation for the feedback field ${C}$. This models a temporal delay,
and the possibility of having asynchronous overlap between pulses that have propagated through different paths.
The output field ${A_{out}}$ = $A_{out}(\tau)$ is not needed for the evolution of the map, but it can be obtained from the coupling
condition at node 2 as
\begin{equation}
A_{out} =  i \sqrt{\theta_2} e^{-i\delta_{1}} A_{m}(L_{1},\tau) + { \sqrt{1 - \theta_2}}e^{-i\delta_{2}}C_{m-1}(L_{2},\tau).
\end{equation}
The details of the homogeneous solutions of the Ikeda map model are described in the Supplementary note 1.

\vspace{10pt}
{*Corresponding author: jboggio2006@gmail.com}

\subsection*{References}

[1] Reed, G. T. Device physics: The optical age of silicon. Nature 427, 595-596 (2004).

[2] Jalali, B. Teaching silicon new tricks. Nat. Photonics 1, 193-195 (2007).

[3] Razzari, L.D. et al. CMOS-compatible integrated optical hyper-parametric oscillator. Nat. Photonics 4, 41-45 (2010).

[4] Levy, J. S. et al. CMOS-compatible multiple-wavelength oscillator for on-chip optical interconnects. Nat. Photonics 4, 37-40 (2010).

[5] Okawachi, Y. et al. Octave-spanning frequency comb generation in a silicon nitride chip. Opt. Lett., 36, 3398-3400 (2011).

[6] Ferdous, F. et al. Spectral line-by-line pulse shaping of on-chip microresonator frequency combs. Nat. Photonics 5, 770–776 (2011).

[7] Kippenberg, T. J., Gaeta, A. L., Lipson, M., Gorodetsky, M. L. Dissipative Kerr solitons in optical microresonators. Science 361, 6402, 2018.

[8] Jost, J. D. et al. Counting the cycles of light using a self-referenced optical microresonator. Optica 2, 706-711 (2015).

[9] Palomo, P. M. et al. Microresonator-based solitons for massively parallel coherent optical communications. Nature 546, 274- 281 (2017).

[10] Dutt, A. et al. On-chip dual-comb source for spectroscopy. Sci. Adv. 4, e1701858 (2018).

[11] Trocha, P. et al. Ultrafast optical ranging using microresonator soliton frequency combs. Science 359, 887–891 (2018).

[12] Kues, M. et al. Quantum optical microcombs. Nat. Photonics 13, 170–179 (2019).

[13] Suh, M.-G. Searching for exoplanets using a microresonator astrocomb. Nat. Photonics 13, 25-30 (2019).

[14] Obrzud, E. et al. A microphotonic astrocomb. Nat. Photonics 13, 31–35 (2019).

[15] Xue, X. et al. Microcomb-based true-time-delay network for microwave beam forming with arbitrary beam pattern control. J. Lightwave Technol. 36, 2312–2321 (2018).

[16] Herr, T. et al. Universal formation dynamics and noise of Kerr-frequency combs in microresonators. Nat. Photonics 6, 480-487 (2012).

[17] Savchenkov, A. et al. Stabilization of a Kerr frequency comb oscillator Opt. Lett. 38 2636-2639 (2013).

[18] Herr, T. et al. Temporal solitons in optical microresonators. Nat. Photonics 8, 145-152 (2014).

[19] Del’Haye, P., Beha, K., Papp, S. B., Diddams, S. A. Self-injection locking and phase-locked states in microresonator-based optical frequency combs. Phys. Rev. Lett. 112, 043905 (2014).

[20] X. Xue, et al. Mode-locked dark pulse Kerr combs in normal-dispersion microresonators. Nat. Photonics 9, 594 (2015).

[21] Del’Haye, P.,  Coillet, A., Loh, W., Beha, K., Papp S. B., and Diddams, S. A. Phase steps and resonator detuning measurements in microresonator frequency combs. Nat. Commun. 6, 5668 (2015).

[22] Brasch, V. et al. Photonic chip based optical frequency comb using soliton Cherenkov radiation. Science 351, 357-360 (2015).

[23] Wang, P.-H. et al. Intracavity characterization of micro-comb generation in the single-soliton regime. Opt. Express 24, 10890-10897 (2016).

[24] Guo, H. et al. Universal dynamics and deterministic switching of dissipative Kerr solitons in optical microresonators, Nat. Photonics 13, 94-102 (2017).

[25] Grudinin, I. S. et al. High-contrast Kerr frequency combs. Optica 4, 434-437 (2017).

[26] Cole, D. C., Lamb, E. S., Del’Haye, P., Diddams S. A., and Papp, S. B. Soliton crystals in Kerr resonators. Nat. Photonics 11, 671-676 (2017).

[27] Yu, M. et al. Breather soliton dynamics in microresonators. Nat. Commun. 8, 14569 (2017).

[28] Pfeiffer, M. H. P. et al. Octave-spanning dissipative Kerr soliton frequency combs in Si$_3$N$_4$ microresonators. Optica 4, 684-691 (2017).

[29] Wang, W. et al. Robust soliton crystals in a thermally controlled microresonator. Opt. Lett. 43, 2002–2005 (2018).

[30] Karpov, M. et al. Dynamics of soliton crystals in optical microresonators. Nat. Phys. 15, 1071–1077 (2019).

[31] Weng, W., Bouchand, R., Lucas, E. et al. Heteronuclear soliton molecules in optical microresonators. Nat Commun 11, 2402 (2020).

[32] Lu, Z. et al. Deterministic generation and switching of dissipative Kerr soliton in a thermally controlled micro-resonator. AIP Advances 9, 025314 (2019).

[33] Lamont, M. R. E., Okawachi, Y., and Gaeta, A. L. Route to stabilized ultrabroadband microresonator-based frequency combs. Opt. Lett. 38, 3478–3481 (2013).

[34] Coen, S. Erkintalo, M. Universal scaling laws of Kerr frequency combs. Opt. Lett. 38 1790-1792 (2013).

[35] Chembo, Y. K., Menyuk, C.R. Spatiotemporal Lugiato-Lefever formalism for Kerr-comb generation in whispering-gallery-mode resonators. Phys. Rev. A 87, 053852 (2013).

[36] Hansson, T., Modotto, D., Wabnitz, S.,  Dynamics of the modulational instability in microresonator frequency combs. Phys. Rev. A, 88 023819 (2013).

[37] Coillet, A. et al. Azimuthal Turing Patterns, Bright and Dark Cavity Solitons in Kerr Combs Generated With Whispering-Gallery-Mode Resonators. IEEE Photonics J. 5, 6100409 (2013).

[38] Hansson, T., and Wabnitz, S.  Bichromatically pumped microresonator frequency combs. Phys. Rev. A 90, 013811 (2014).

[39] Godey, C., Balakireva, I. V., Coillet, A., and Chembo, Y. K. Stability analysis of the spatiotemporal Lugiato-Lefever model for Kerr optical frequency combs in the anomalous and normal dispersion regimes. Phys. Rev. A 89, 063814 (2014).

[40] Coen S. and Haelterman, M. Impedance-matched modulational instability laser for background-free pulse train generation in the THz range. Optics Communications, 146, 339-346, 1998.

[41] Xue, X., Wang, P.-H., Xuan, Y., Qi, M., and Weiner, A. M. Microresonator Kerr frequency combs with high conversion efficiency. Laser \& Photon. Rev. 11, 1600276 (2017).

[42] Miller, S. A. et al. Tunable frequency combs based on dual microring resonators. Optics Express, 23, 21527-21540 (2015).

[43] Xue, X. et al. Normal-dispersion microcombs enabled by controllable mode interactions. Laser \& Photon. Rev. 9, 23-28 (2015).

[44] Kim, S., Han, K., Wang, C. et al. Dispersion engineering and frequency comb generation in thin silicon nitride concentric microresonators. Nat Commun 8, 372 (2017).

[45] Xue, X., Zheng, X. and Zhou, B. Super-efficient temporal solitons in mutually coupled optical cavities. Nat. Photonics 13, 616–622 (2019).

[46] Jang, J. K. et al. Observation of Arnold Tongues in Coupled Soliton Kerr Frequency Combs, Phys. Rev. Lett. 123, 153901 (2019).

[47] Helgason, Ó.B., Arteaga-Sierra, F.R., Ye, Z. et al. Dissipative solitons in photonic molecules. Nat. Photonics 15, 305–310 (2021).

[48] Tikan, A., Riemensberger, J., Komagata, K. et al. Emergent nonlinear phenomena in a driven dissipative photonic dimer. Nat. Phys. 17, 604–610 (2021).

[49] Yariv, A. Critical coupling and its control in optical waveguide-ring resonator systems. IEEE Photonics Technology Letters, 14, 483-485, 2002.

[50] Green, W. M. J., Lee, R. K., DeRose, G. A., Scherer, A., and Yariv, A. Hybrid InGaAsP-InP Mach-Zehnder Racetrack Resonator for Thermooptic Switching and Coupling Control. Opt. Express 13, 1651-1659 (2005).

[51] Zhou L. and Poon, A. W. Electrically reconfigurable silicon microring resonator-based filter with waveguide-coupled feedback. Opt. Express 15, 9194-9204 (2007).

[52] Chen, L., Sherwood-Droz, N., and Lipson, M. Compact bandwidth-tunable microring resonators. Opt. Lett. 32, 3361–3363 (2007).

[53] Matsko, A. B., Savchenkov, A. A., Strekalov, D., Ilchenko, V. S., and Maleki, L.  Optical hyperparametric oscillations in a whispering-gallery-mode resonator: Threshold and phase diffusion. Phys. Rev. A 71, 033804 (2005).

[54] Pasquazi, A. et al. Stable, dual mode, high repetition rate mode-locked laser based on a microring resonator. Opt. Express 20, 27355–27362 (2012).

[55] Johnson, A.R., Okawachi, Y., Lamont, M. R. E., Levy, J. S., Lipson, M. and Gaeta, A. L.  Microresonator-based comb generation without an external laser source. Opt. Express 22, 1394-1401 (2014).

[56] Bao, H., et al. Laser cavity-soliton microcombs. Nat. Photonics 13, 384–389 (2019).

[57] Kues, M., Brauckmann, N., Walbaum, T., Gross, P., Fallnich, C. Nonlinear dynamics of femtosecond supercontinuum generation with feedback. Opt Express 17, 15827-15841 (2009).

[58] Brauckmann N, Kues M, Walbaum T, Gross P, Fallnich C. Experimental investigations on nonlinear dynamics in supercontinuum generation with feedback. Opt Express 18, 7190-7202 (2010).

\newpage

\setcounter{figure}{0}
\renewcommand\thefigure{S\arabic{figure}}

\section*{Supplementary Information for:
Efficient Kerr soliton comb generation in micro-resonators with interferometric back-coupling}
\vspace{2mm}

J.M. Chavez Boggio, D. Bodenmüller, S. Ahmed, S. Wabnitz, D. Modotto, and T. Hansson

\subsection*{Supplementary note 1: Analysis of homogeneous solutions, detuning relations and resonance width}
\vspace{2mm}

\begin{figure}[h]
\centering
\includegraphics[width=0.65\linewidth]{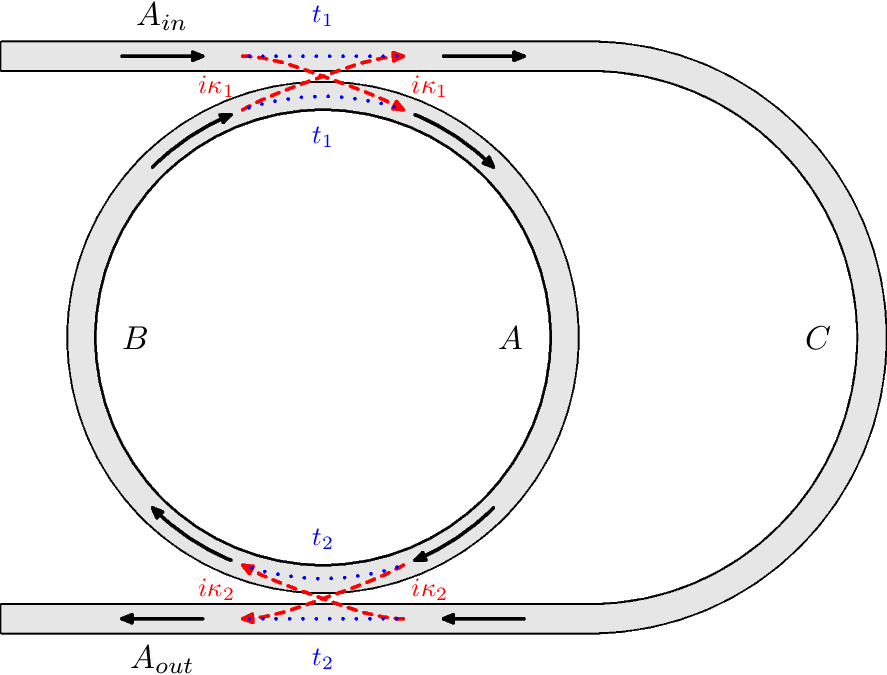}
\caption{Schematic of the microresonator device with optical feedback. The fields as well as the transmission and coupling coefficients are indicated at various locations.}
\label{figure:S2}
\end{figure}

The homogeneous solutions of the Ikeda map are found by considering the pump frequency only, and disregarding dispersive terms.
The solution for propagation in each waveguide section is given by
\begin{equation}
 \begin{aligned}
A_{m}(L_{1}) = A_{m}(0) \exp\left[-\dfrac {\alpha_i L_{1}}{2} + i \gamma L_{eff}(L_{1}) |A_{m}(0)|^{2}\right]
 \end{aligned}
\end{equation}
where ${L_{eff}(L_1)} = {(1 - e^{-\alpha_i L_{1}})}/{\alpha_i}$, with analogous formulas for the fields ${B_{m}}(L_1)$ and ${C_{m}}(L_2)$. In the stationary case we must require the fields to
reproduce themselves after each roundtrip. The roundtrip index can then be dropped, to obtain
\begin{equation}
 \begin{aligned}
{A(0)} = i{\sqrt{\theta_1}}A_{in} + {\sqrt{1 - \theta_1}}e^{-i\delta_{1}}B(L_{1}), \qquad {C(0)} = {\sqrt{1 - \theta_1}}A_{in} + i{\sqrt{\theta_1}}e^{-i\delta_{1}}B(L_{1}),
 \end{aligned}
\end{equation}
for the coupling conditions at node 1, and
\begin{equation}
 \begin{aligned}
{B(0)} = {\sqrt{1 - \theta_2}}e^{-i\delta_{1}}A(L_{1}) + i{\sqrt{\theta_2}}e^{-i\delta_{2}}C(L_{2})
 \end{aligned}
\end{equation}
for the coupling condition at node 2.
By introducing the shorthand notation
${t_{j} = \sqrt{1 - \theta_j}}$ and ${\kappa_{j} = \sqrt{\theta_j}}$,\, and the power dependent functions
\begin{equation}
  \begin{aligned}
{a} = t_{1} - t_{2} g_{A} g_{B}, \qquad {c} = \kappa_{1} + \kappa_{2} g_{B} g_{C},
  \end{aligned}
\end{equation}
where  ${g_k }$ = ${\exp[-{(\alpha_i/{2)} L_j} - {i\delta_j} + i\gamma L_{eff} (L_j)I_k]}$ with ${j}$ = 1 if ${k}$ = ${A},{B}$ and ${j}$ = 2 if ${k}$ = ${C}$. The coupling conditions can be rewritten as equations for the fields at $z = 0$ as
\begin{equation}
 \begin{aligned}
 {A(t_1a + \kappa_1c)} = ic A_{in}, \qquad {C(t_1a + \kappa_1c)} = a A_{in},
 \end{aligned}
\end{equation}
\begin{equation}
 \begin{aligned}
 {g_B B(t_1a + \kappa_1c)} = i(t_1c - \kappa_1a) A_{in}, \qquad {A_{out}(t_1a + \kappa_1c)} = (t_2ag_C - \kappa_1cg_A) A_{in}.
 \end{aligned}
\end{equation}
These equations can be used to solve for the powers, ${I_A}$ = ${|A|^{2}}$, ${I_B}$ = ${|B|^{2}}$, 
${I_C}$ = ${|C|^{2}}$, ${I_{in}}$ = ${|A_{in}|^{2}}$ and ${I_{out}}$ = ${|A_{out}|^2}$, so that one obtains the following
closed system

\begin{equation}
 \begin{aligned}
{I_A = \frac{|c|^{2}}{|t_1a + \kappa_1c|^{2}} I_{in}}, \qquad {I_C = \frac{|a|^{2}}{|t_1a + \kappa_1c|^{2}} I_{in}}, \qquad {I_B = \frac{|t_1c - \kappa_1a|^{2}}{|g_B|^{2}|t_1a + \kappa_1c|^{2}} I_{in}}. \label{eq:pwr}
 \end{aligned}
\end{equation}
These powers are related through the conditions ${I_{in}}$ + ${|g_B|^{2}I_B}$ = ${I_A}$ + ${I_C}$, $I_{out} + I_B = |g_A|^2I_A+|g_C|^2I_C$, and directly give the solution in the linear case (i.e., for $\gamma \to 0$).

In order to retrieve the detuning relations, we consider the phase shifts acquired by the field during propagation through section A and C, which are given by
\begin{equation}
\begin{aligned}
{\phi_1} = \beta L_{1} \approx (\beta_0 + \beta_1 \Delta \omega)L_{1} = 2\pi {m_1} - \delta_{1},
\end{aligned}
\end{equation}
\begin{equation}
 \begin{aligned}
{\phi_2} = \beta L_{2} \approx (\beta_0 + \beta_1 \Delta \omega) (3L_{1} + \Delta) = 2\pi {m_2} - \delta_{2},
 \end{aligned}
\end{equation}
where ${\beta = \beta(\omega)}$ is the propagation constant, $\beta_0 = \beta(\omega_0)$ and ${\Delta}$ is the tunable difference in length which is provided by the feedback arm. We have that ${\beta_0L_1}$ = $2\pi m_1$, ${3\beta_0L_1}$ = $2\pi m_2$, and $m_2 = 3m_1$. The detuning is related to the angular frequency difference $\Delta\omega$ between the pump and the resonance frequency $\omega_0$ through
\begin{equation}
 \begin{aligned}
{\delta_1} = -\beta_{1}L_{1} \Delta \omega,
 \end{aligned}
\end{equation}
\begin{equation}
 \begin{aligned}
{\delta_2} = -3\beta_{1}L_{1} \Delta \omega - \beta_{0} {\Delta} - \beta_{1} {(\Delta \omega)} {\Delta} \cong {3\delta_1} - {\beta_0 \Delta} = 3\delta_1 + \delta_{20},
 \end{aligned}
\end{equation}
where $\delta_{20} = -\beta_0\Delta$ is a detuning offset. The propagation time in each section is given by ${t_1 = \beta_{1}L_{1}}$ and $t_2 = \beta_{1}(3L_{1} + \Delta) = 3t_1 + \beta_{1} \Delta$. With the latter
being related to the walk-off through ${3L_{1}(\Delta \beta_{1})}$ = ${\beta_{1} \Delta}$.

The resonance width can be determined from the expression for $I_B$ in Eq. (\ref{eq:pwr}). The resonance occurs when the imaginary part of the denominator is zero, while the width is obtained by setting the imaginary part equal to the real part. The resonance width is found to be approximately given by 
 \begin{equation}
  \begin{aligned}
\Delta\omega_{T} =  [\alpha L_1 + (\theta_1 + \theta_2)/2  + \sqrt{\theta_1\theta_2} \cos{(\beta_0 \Delta)}]/ \beta_1 L_1,
 \end{aligned}
\end{equation}
where ${\alpha_i L_1}$ is the absorption loss contribution to the linewidth, while the second and third terms account for the contribution of the coupling coefficients. By adjusting $\Delta$, the coupling condition is tuned, which changes the linewidth of the resonance, and also changes the loaded \emph{Q}-factor [1,2].

 \subsection*{Supplementary note 2: Resonance characterization}
\vspace{1mm}

To characterize the transmission properties of the resonators with feedback used in this paper to generate DKSs, a frequency-sweeping interferometric technique is employed\, [3]. This technique allows us to measure the resonance linewidth and depth, the\, FSR dispersion, and\, \emph{Q}-factors as a function of wavelength. The light of a wavelength-tunable laser with 5\, mW\, CW power is swept from 1550 to 1630\, nm with a 10\, nm/s speed. The light beam is split in two, 50 \% of it is used to scan the resonances of the resonator with feedback, and the other 50\% is injected into a Mach-Zehnder interferometer for generating a low frequency ruler with $\sim 20$\, MHz periodicity. Any irregularity of the laser scan velocity, when measuring the resonances, is tracked by the frequency ruler [4]. The transmitted spectrum through the resonator with feedback and the frequency ruler are both visualized with an oscilloscope at a frequency sampling rate of one point/MHz.

\begin{figure}[h]
\centering
\includegraphics[width=\linewidth]{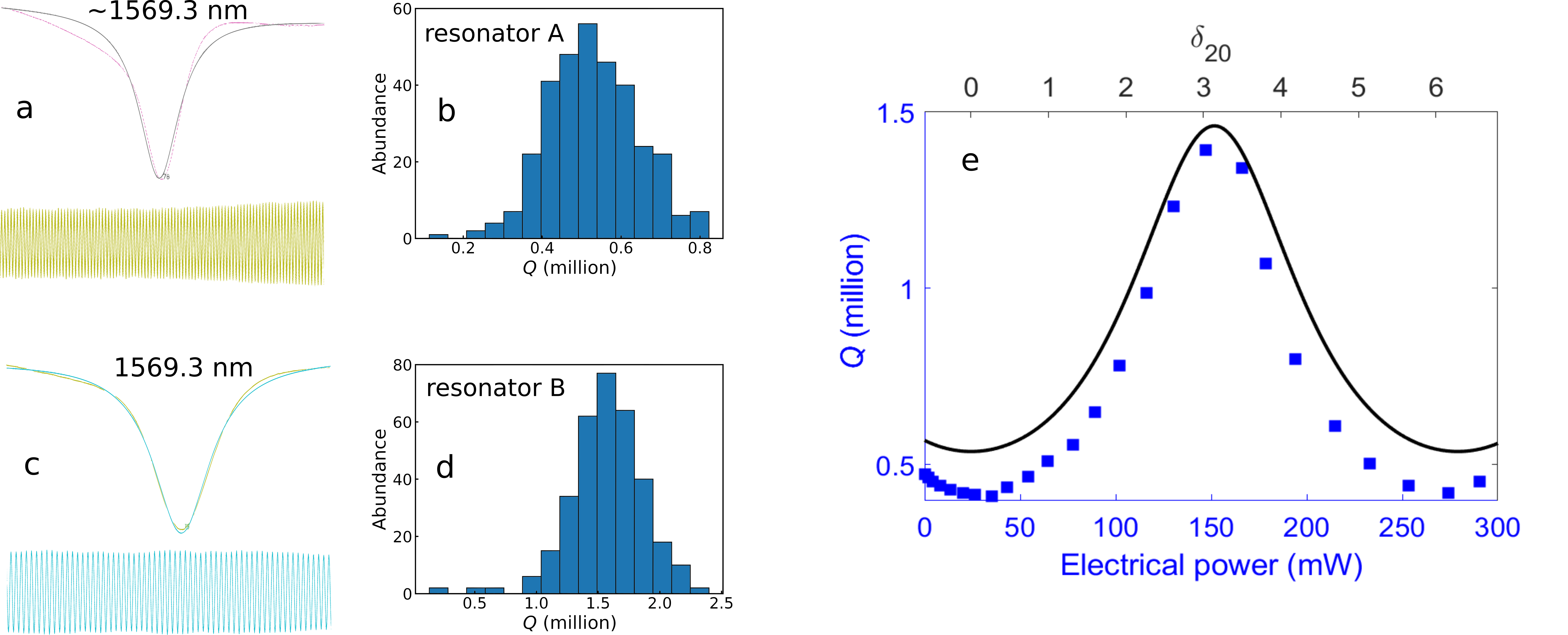}
\caption{Resonance characterization of (a,b) resonator A and (c,d) resonator B. (a,c) Zoom of resonance at 1569.3 nm. Both resonators have $R = 800\, \mu$m, and first gap between bus waveguide and ring of gap1 = 650\, nm, but the second gap is 410 nm for resonator A and 460 nm for resonator B. (b,d) corresponding \emph{Q}-factor abundances by scanning the resonances from 1550 to 1630 nm. (e) Interferometric tuning of the coupling condition in resonator A. ${\Delta}$\, (${\delta_{20}}$) is changed by applying electrical power on a thermal heater. The  measured average \emph{Q}-factor for resonator A as a function of the electrical power is depicted in blue squares, while the calculated values from equation (18) are shown in a solid black line.}
\label{figure:S1}
\end{figure}

Figure S2(a) shows a zoom-in of the resonance pumped to generate the frequency comb spectra in Figures 2(e-h) and Figures 5(a-b) (i.e. resonator A), respectively. Note that the resonance does not exhibit a Lorentzian shape, but is distorted owing to the interferometric coupling. The ruler having a $\sim 20$\,MHz periodicity is shown on the bottom. Figure S2(c)\, shows the zoom-in of the resonance pumped to generate the frequency comb power traces in Figures 4(a-d), and the frequency comb spectra in Figures 5(e-j)\, (i.e. resonator B). The resonances in Figures S2(a) and S2(c) are fitted in order to extract their widths and extinction ratios, and to retrieve the loaded \emph{Q}-factor.
Figures S2(b) and S2(d) show the loaded \emph{Q}-factor abundance from the 330 resonances scanned from 1530 to 1630 nm for the resonators A and B, respectively.

To better characterize the intrinsic losses in our resonators, we measured the loaded \,\emph{Q}-factor for several contiguous standard resonators fabricated with ${R = 132\mu}$m (FSR = 172\, GHz) and gaps covering 500 - 600\, nm, resulting in a measured average\, loaded \emph{Q}-factor in the range ${2.2-2.4\times10^{6}}$. Those extra standard resonators can be observed in the picture shown in the inset of Figure 2(a) in the main text.
 
Figure S2(e) shows, for resonator A, the measured average Q-factor variation as a function of the electrical power in the micro-heater (blue squares), as well as the calculated values (black line). As can be seen, the agreement between measurements and theory is very good. The dependence of the average \emph{Q}-factor on  ${\Delta}$\, follows a very similar behaviour to that of the other resonator with feedback (resonator B), which is shown in Figure 6(a) in the main text: this testifies of the high fabrication uniformity.

\subsection*{Supplementary note 3: Conversion efficiency vs number of propagating solitons}
\vspace{2mm}

Very high conversion efficiencies were obtained for soliton crystals having around 50 pulses circulating in the ring of a microresonator with optical feedback. On the other hand, in isolated resonators it is not possible to reach high conversion efficiencies even when 50 pulses propagate in the resonator. Although we could not experimentally generate a small number of solitons, due to the thermo-optic effect in silicon nitride, we describe here by numerical simulations how the conversion efficiency varies with the number of circulating solitons. For the same parameters as in Figure 3 (${\delta_1}$ = 0.033), we propagated a varying number of pulses in a resonator with feedback. After 10000 round trips, the pulses have evolved into DKSs, in all cases. Figures S3(a,b) and (c,d) show the cases for 10 and 1 solitons, respectively. The spectra at the output (top), ring (middle), and feedback (bottom) sections show the evolution of the frequency comb for both cases. While the conversion efficiency is reasonably good for the 10-soliton case (around 7\%), for the single-soliton case the efficiency is only below 1\%.

\begin{figure}[h]
\centering
\includegraphics[width=0.9\linewidth]{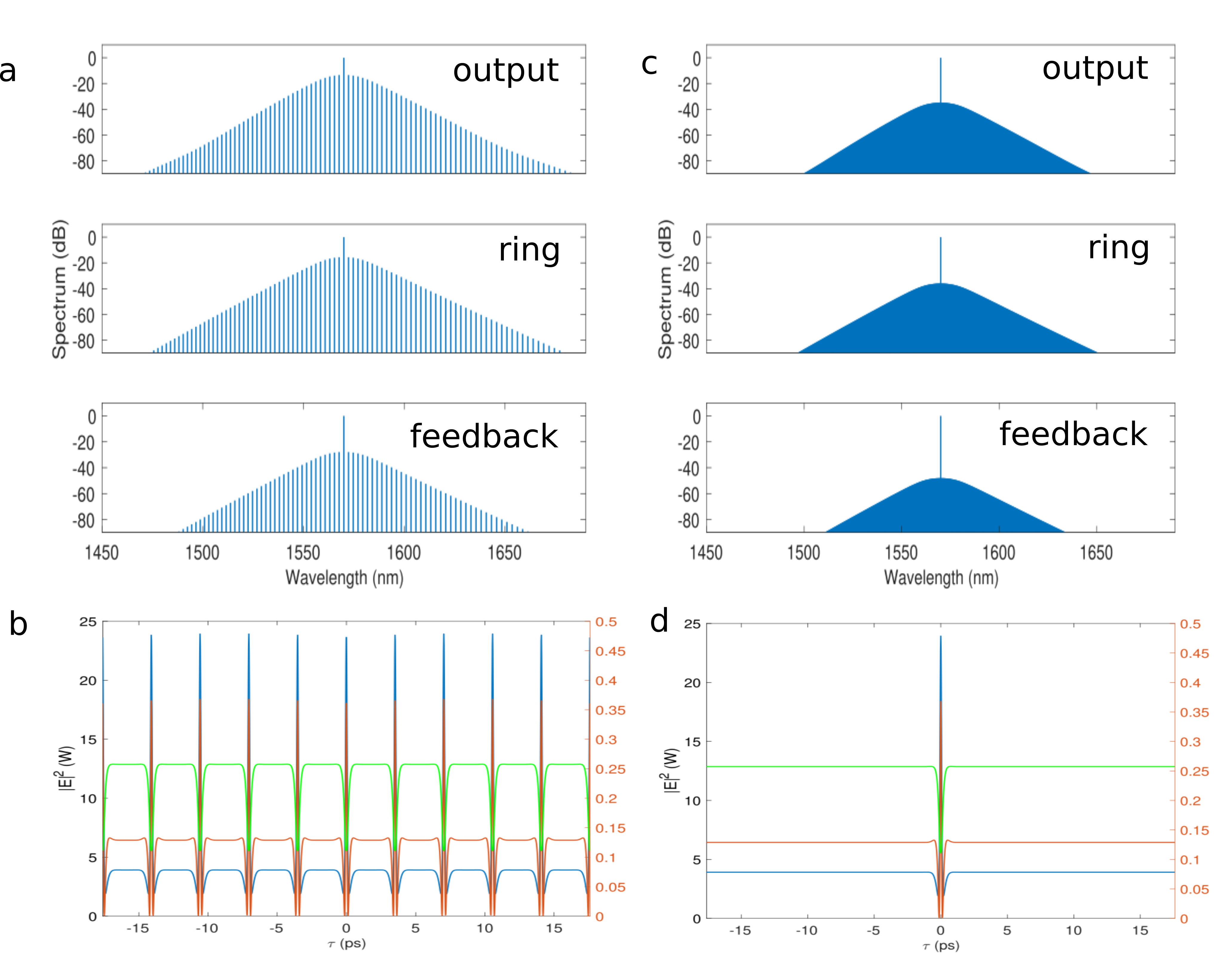}
\caption{ Frequency comb spectra after propagation over 10000 round-trips of (a) ten equidistant pulses (c) one pulse. The spectra are calculated at the output (top), ring (middle), and feedback (bottom) sections.
The corresponding temporal fields are shown in (b) and (d): output (red), ring (blue), and feedback (green) sections. }
\label{figure:S3}
\end{figure}

The conversion of the input pump power into the output power of comb lines is limited by the amount of pump power that remains circulating in the ring, in order to sustain the solitons. Simulations were performed by keeping ${\theta_1}=$ ${4.1\times10^{-3}}$, while increasing ${\theta_2}$ from ${2.38\times10^{-2}}$ up to ${3.77\times10^{-2}}$, and finally to ${8\times10^{-2}}$. The amount of pump power that circulates in the ring goes from 75\%, to 49\% and finally it drops down to 25\%, showing that the conversion efficiency into the output comb lines power can be made extremely large.   

\subsection*{Supplementary note 4: Accessing single solitons}
\vspace{1mm}

Multi-soliton steps were measured routinely in microresonators with optical feedback, as shown in Figure 4(a-d) of the main text. However, due to thermal effects, we could not access those states by direct tuning of the pump laser. Nevertheless, we could measure effects associated to the propagation of intense coherent pulses circulating in the ring: Raman self-frequency shift and third harmonic generation. Figure S4(a)\, shows the frequency comb spectrum obtained by pumping with 100\, mW a resonator with feedback with the following parameters: gap1 = 650 nm, gap2 = 440 nm, FSR = 172 GHz, and loaded \emph{Q}-factor of ${10^6}$. As can be seen, there is a small but noticeable Raman shift. The pump is only 17 dB stronger than the strongest comb line, indicating a moderate conversion efficiency. Figure S4(b)\, shows the frequency comb spectrum obtained by pumping with 400\, mW a resonator with feedback having gap1 = 600 nm,\, gap2 = 440 nm,\, FSR = 172 GHz,\, and\, loaded \emph{Q}-factor of ${10^6}$. The Raman shift can be clearly seen and it is around 1.4 THz, indicating the generation of intense pulses. Furthermore, the inset shows the picture of the chip taken during the frequency comb generation of the spectrum in Figure S4(b). It can be noted the scattering of strong green light, due to third harmonic generation in the ring resonator. This indicates that even though the frequency comb spectral shape in Figure S4(b) does not exhibit a perfect sech${^2}$ profile, which is expected for pure DKS generation, a very intense and coherent pulse is circulating inside the ring, which gives rise to the observed Raman shift and third-harmonic green light.

\begin{figure}[h]
\centering
\includegraphics[width=\linewidth]{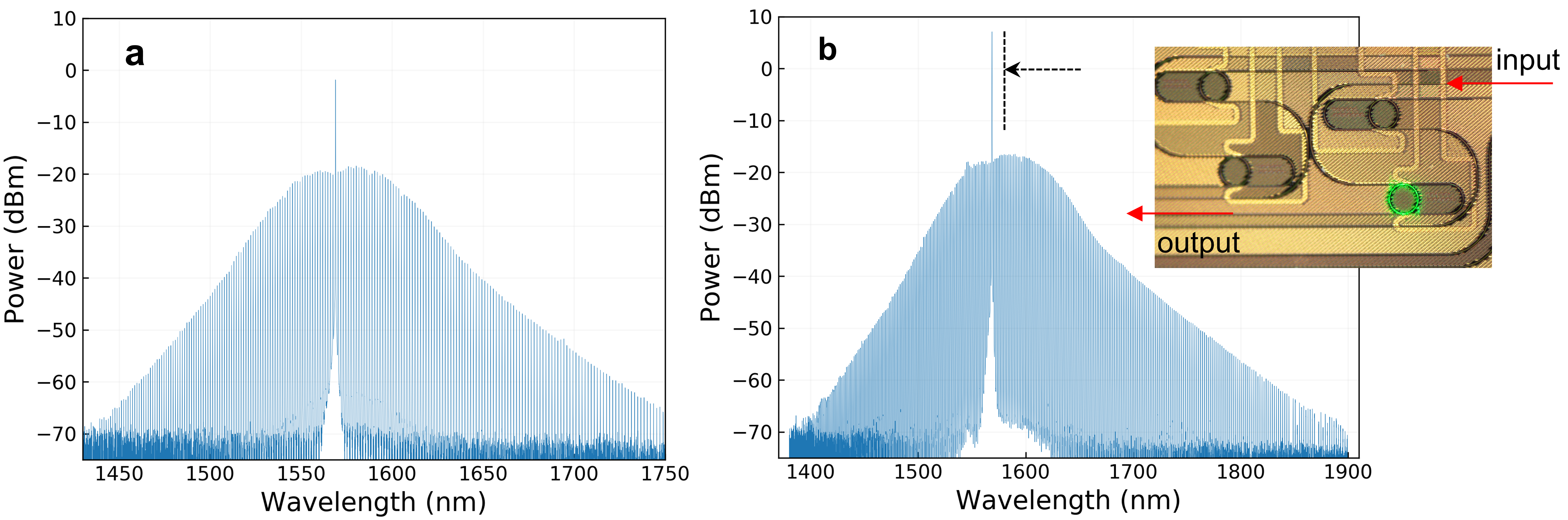}
\caption{ Frequency comb spectra for: (a) 100\, mW and (b) 400\, mW. Inset: Picture of the chip showing green light generation due to a third-harmonic generation process originating from the intense pulses associated with the frequency comb in (b).}
\label{figure:S4}
\end{figure}

\subsection*{Supplementary note 5: Power of the generated comb as a function of detuning}
\vspace{1mm}

Figure S5 shows one of the traces of the generated comb power as a function of pump detuning from resonance as depicted in Figure 4(b), which was obtained with a pump power of 68 mW. The frequency comb spectra for three different detuning values are also shown, at points which are indicated with arrows. At low detuning, a Turing roll is generated; by increasing the value of the detuning, a modulation instability spectrum is obtained. By further increasing the pump detuning, a PSC is finally generated. The difference between the spectrum of a Turing roll and that of a PSC lies in the separation between the comb lines and the bandwidth, which is broader for the PSC.

\begin{figure}[h]
\centering
  \includegraphics[width=0.8\linewidth]{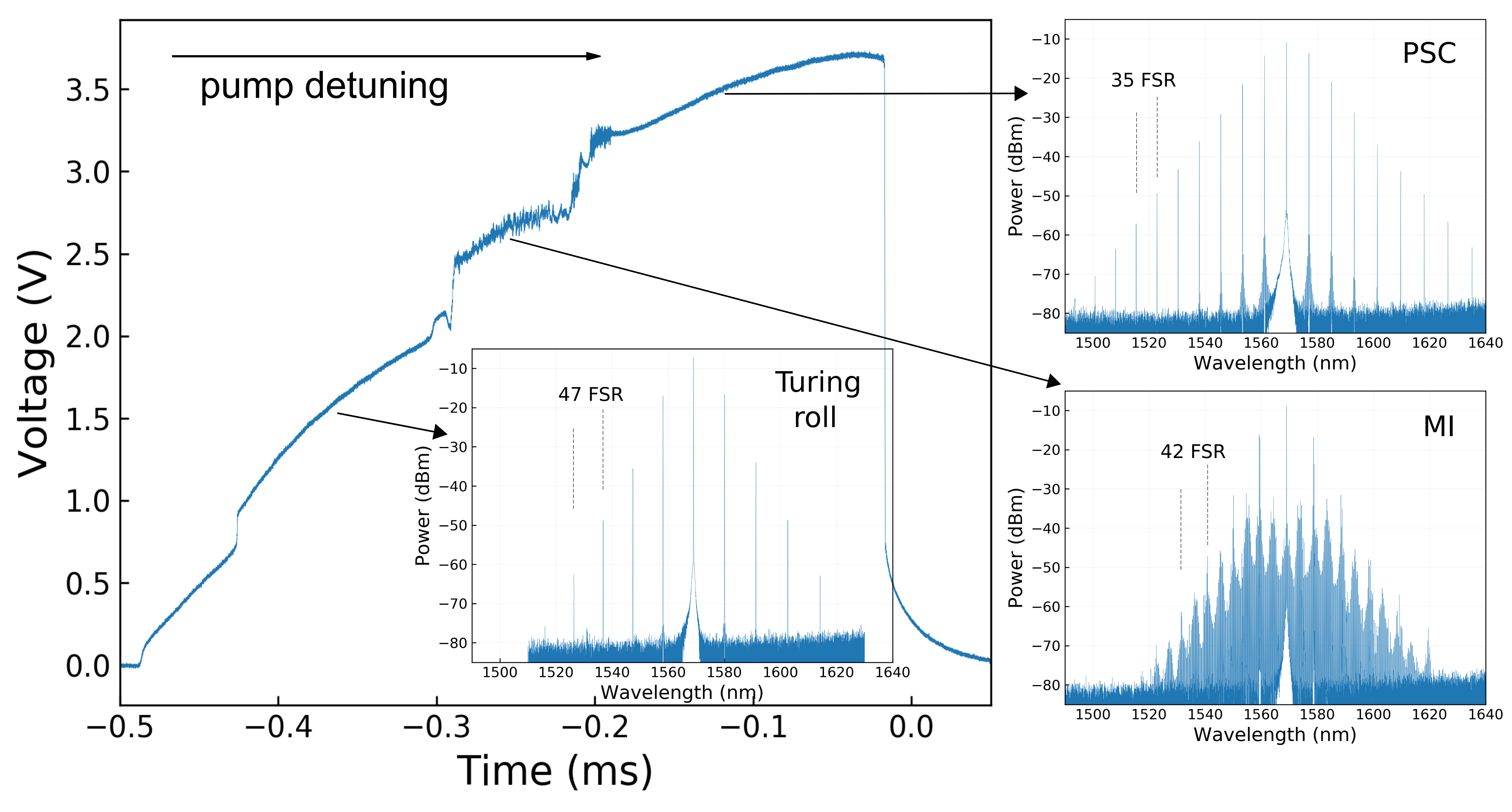}
\caption{Generated comb light as a function of pump-resonance detuning in a resonator with feedback from Figure 4(b) in the main text. The frequency comb spectra generated at different detuning values are indicated by an arrow.}
\label{figure:S5}
\end{figure}

\subsection*{Supplementary note 6: Parametric threshold and PSC conversion efficiency vs pump power}
\vspace{5mm}

From the various measurements performed with a resonator with feedback, a trend was observed: by keeping the phase mismatch between the main and the feedback cavities at a constant value, but varying the input pump power, the conversion efficiency could be changed. This observation is illustrated in Figures S6(b-e), where we plot the spectra of the generated perfect SCs for pump powers of 20, 30, 40, and 50 mW, respectively. These results come from the same set of measurements shown in Figure 4(a-d) in the main text. Note that, by reducing the pump power from 50 to 30 mW, the pump is progressively depleted due to the enhancement of the conversion efficiency. For a pump power of 20 mW this trend is slightly reversed. For a comparison, Figures S6(f) shows the frequency comb spectrum for a pump power of 40 mW, but for a pump detuning that generates a Turing roll: in this case the conversion efficiency is much smaller.

Interestingly, the threshold for parametric generation, i.e., for the appearance of the first comb lines, is at\, $\sim15$\, mW, as shown in Figure S6(a). This is slightly smaller than the expected value of 20.5 mW for a single resonator (i.e. for $f = 1$), and is at the edge of the experimental error in determining the pump power inside the chip.

\begin{figure}[h]
\centering
\includegraphics[width=0.8\linewidth]{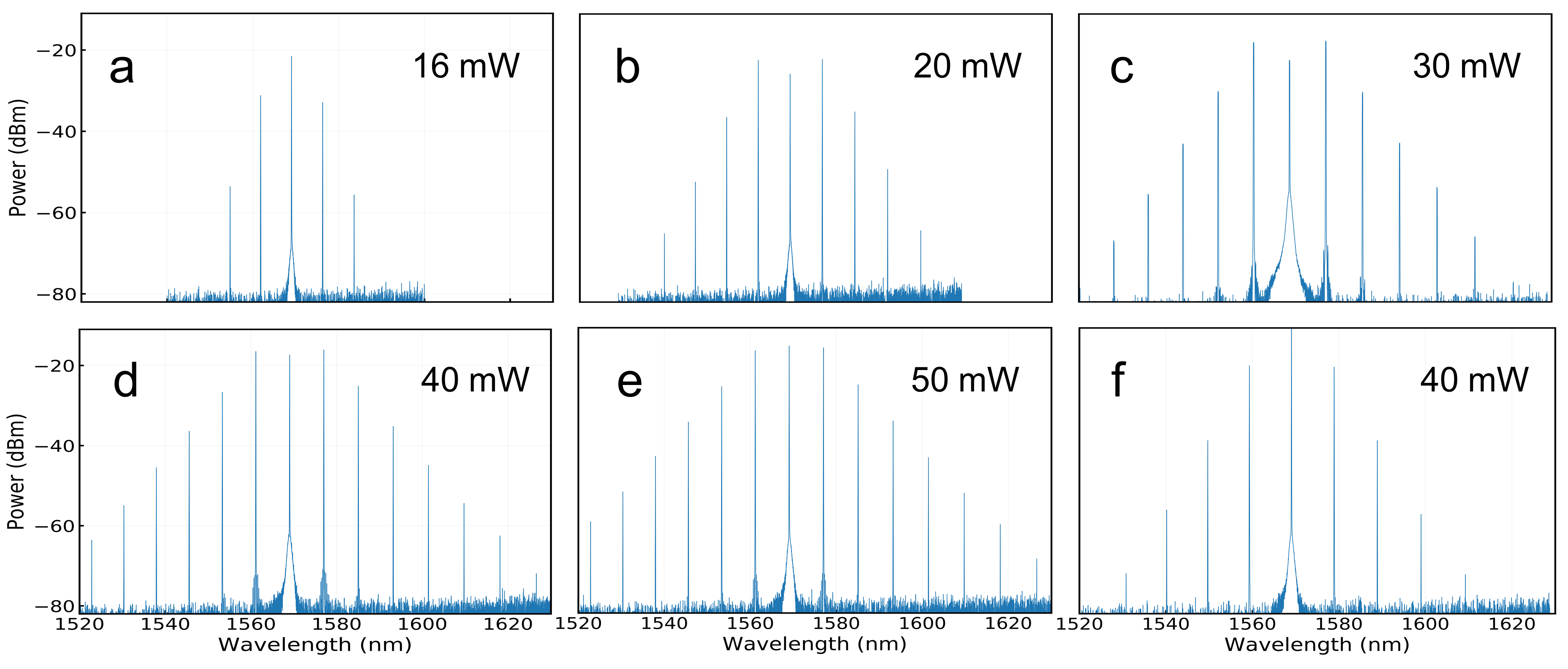}
\caption{(a-e) Generated frequency comb spectra for pump powers of\, 16,\, 20,\, 30, 40,\, and 50\, mW, respectively. (f) shows the Turing roll generated for 40 mW of pump power.}
\label{figure:S6}
\end{figure}

\subsection*{Supplementary note 7: Stability of the repetition rate}
\vspace{5mm}

The coherence of frequency combs generated in our resonators with feedback was assessed through repetition rate measurements. The pump comb line was filtered out, and the whole comb spectrum was detected with a high-speed photodiode. The stability of the repetition rate was measured with a frequency counter, after performing a down-conversion from $28.4$\, GHz to sub-100\, MHz. The frequency of the down-converted signal was measured with a frequency counter at different gate times. Every 5\, minutes the gate time was changed. Figures S7(a-e) show the variation of the repetition rate over a time span of 30 minutes. A general trend that can be observed is a frequency drift towards smaller frequencies. Furthermore, jumps of the repetition rate frequency can be noted at $t = 755$ and at $t = 845$ seconds. Those jumps can be better observed for a gate time of 100\, ms, when fast fluctuations are averaged out. Note that the fast fluctuations are a bit weaker at the beginning of the frequency comb operation. Interestingly, although the repetition rate signal of the frequency comb was always intense, the spectrum changed in a strong way. Figure S7(f) shows the spectrum one minute after the frequency comb is generated, while Figure S7(g) shows the OFC spectrum 15 minutes later. The strong lines have almost disappeared, indicating that a number of DKSs were annihilated. 
The comb spectrum in Figure S7(h) was taken ten minutes later: it exhibits a small change with respect to S7(g). Since our setup was not enclosed, it was subject to environmental disturbances, such as air flows, making it susceptible to drifts of the in-coupled pump power. Nevertheless, even though no stabilisation was performed, the robustness of soliton operation in the resonator with extended cavity turned out to be truly remarkable.

\begin{figure}[h]
\centering
\includegraphics[width=0.5\linewidth]{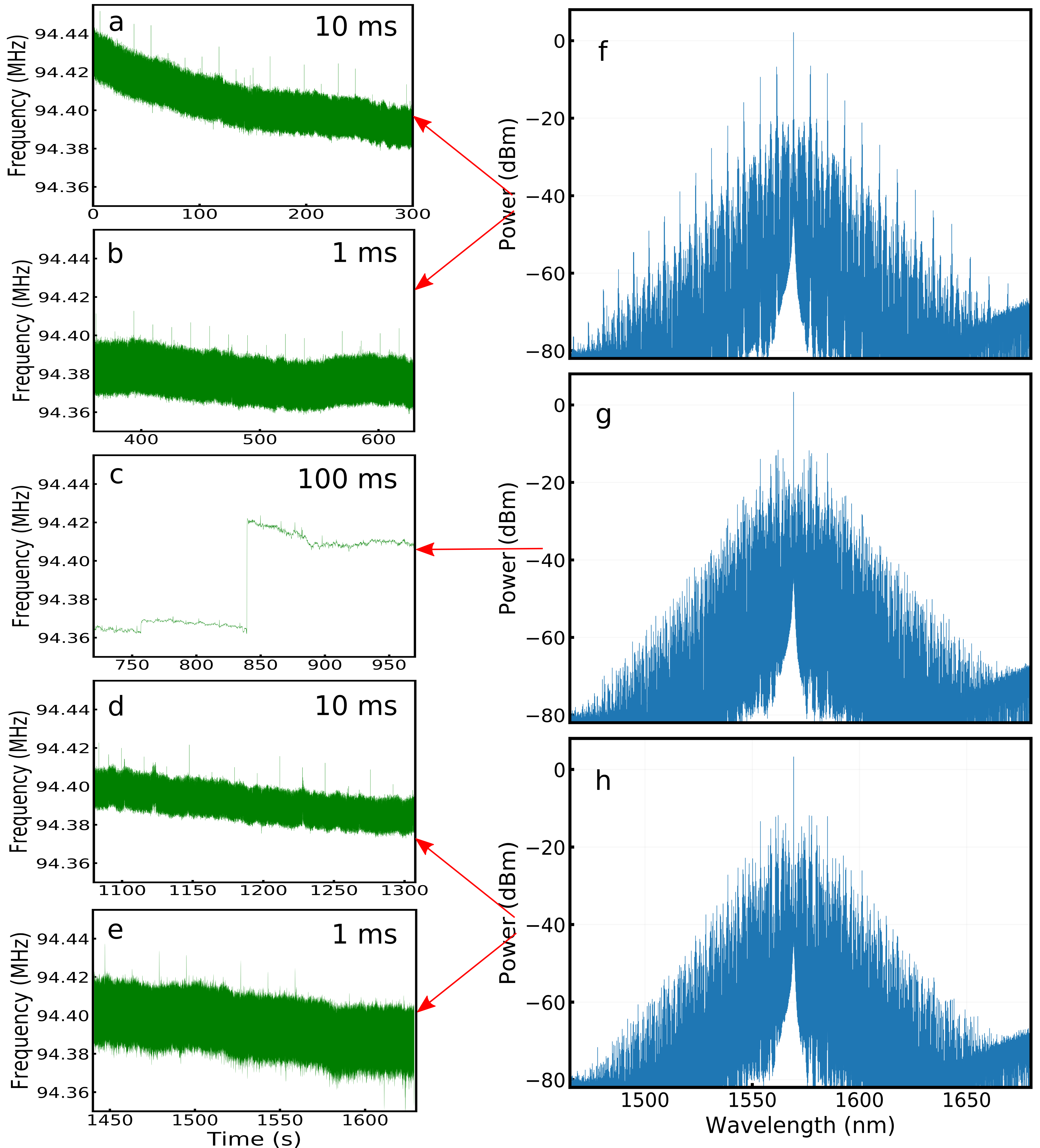}
\caption{(a-e) Repetition rate stability of the soliton crystal with defect measured with a frequency counter over 30 minutes. (f-h) Frequency comb spectral changes in time. Although high coherence was preserved, the spectrum changed, which might be due to soliton decays.
}
\label{figure:7}
\end{figure}

\subsection*{Supplementary note 8: Inverse Fourier transform of experimental spectra}
\vspace{5mm}

Frequency comb spectra reported in the main text were red-detuned and exhibit very intense repetition rate signals. Therefore, phase-locked comb lines and soliton formation are expected. Although we did not measure the phases of the comb lines or the autocorrelation traces in order to characterize the generated temporal profiles, the reported spectra show features in the small comb lines that are very similar to those which were reported in the literature on soliton crystals. Because of this, we may assume as a working hypothesis that the strong comb lines are all in phase, while the weak ones are out of phase with respect to the strong ones, i.e. with a phase difference of ${\pi}$ [5,6]. Figures S8(a-d) show four frequency comb output spectra (left), along with their corresponding inverse Fourier transforms (right). The case of Figure S8(a) shows a perfect soliton crystal with moderate conversion efficiency, where its inverse Fourier transform shows 34 equidistant pulses. Note that there is no pedestal in the output pulses: this is related to the fact that the residual pump component has neither too large nor too small intensity, but exactly the appropriate one in order to cancel the background. For Figure S8(b) the inverse Fourier transform shows that a pulse is much weaker than all other 56 pulses, thus creating a vacancy. On the other hand, for Figures S8(c,d) the inverse Fourier transforms show that some pulses are weaker than all other pulses, so that the defects of these soliton crystals are more complex than the single vacancy case. 

\begin{figure}[h]
\centering
\includegraphics[width=\linewidth]{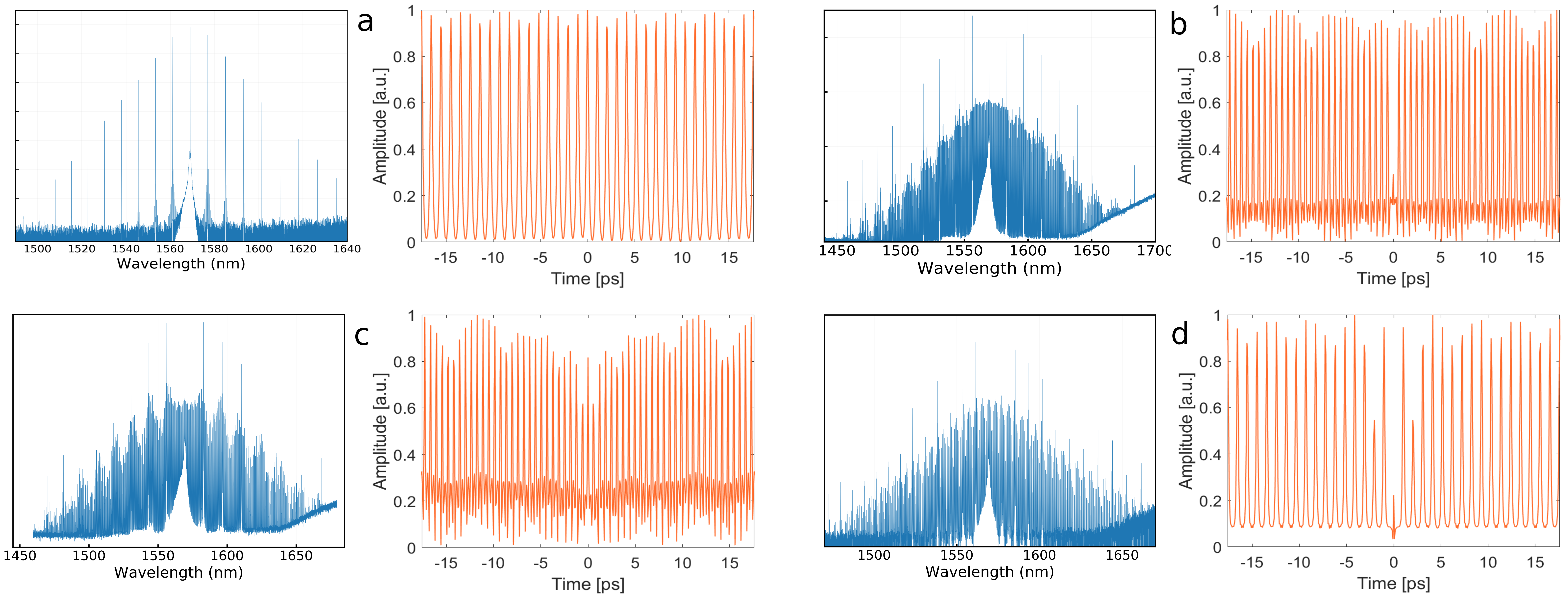}
\caption{(a-d) Frequency comb spectra (left) with their calculated inverse Fourier transform (right) whose period is 35.2 ps.}
\label{figure:S7}
\end{figure}

\subsection*{References}
\vspace{5mm}

[1] Yariv, A. Critical coupling and its control in optical waveguide-ring resonator systems. IEEE Photonics Technology Letters 14, 483-485 (2002).

[2] Chen, L., Sherwood-Droz, N., and Lipson, M. Compact bandwidth-tunable microring resonators. Opt. Lett. 32, 3361–3363 (2007).

[3] Pereira Cabral, A., Rebordão, J.M. Accuracy of frequency-sweeping interferometry for absolute distance metrology. Opt. Eng. 46 073602 (2007).

[4] Del'Haye, P., Arcizet, O., Gorodetsky, M. L., Holzwarth, R., and Kippenberg, T. J. Frequency comb assisted diode laser spectroscopy for measurement of microcavity dispersion. Nature Photon. 3, 529 (2009).

[5] Del’Haye, P.,  Coillet, A., Loh, W., Beha, K., Papp S. B., and Diddams, S. A. Phase steps and resonator detuning measurements in microresonator frequency combs. Nat. Commun. 6, 5668 (2015).

[6] Brasch, V. et al. Photonic chip based optical frequency comb using soliton Cherenkov radiation. Science 351, 357-360 (2015).

\end{document}